\documentclass[letterpaper,12pt]{article}
\usepackage[latin1]{inputenc}
\usepackage{amsmath, graphicx, layout, verbatim,rotating}
\usepackage{setspace}
\usepackage{fancyhdr}
\usepackage{float}
\usepackage{color}
\usepackage{multirow}
\usepackage{natbib}
\usepackage{verbatim}
\usepackage{subfig}
\usepackage{amsthm}
\usepackage{amsfonts}
\usepackage{algorithm}
\usepackage{algpseudocode}
\usepackage{amssymb}

  \usepackage{enumitem}
\setlist{nolistsep}

\newcommand{\Lim}[1]{\raisebox{0.5ex}{\scalebox{0.8}{$\displaystyle \lim_{#1}\;$}}}

\newcommand{\BigO}[1]{\ensuremath{\mathcal{O}\bigl(#1\bigr)}}

\renewcommand{\algorithmicrequire}{\textbf{\small Input:}}
\renewcommand{\algorithmicensure}{\textbf{\small Output:}}

\usepackage{setspace}

\def\X{***ATT***}

\renewcommand{\vec}[1]{\mathbf{#1}}
\renewcommand{\comment}[1]{}

\def\I{{\mathbb I}}

\def\E{{\mathbb E}}
\def\V{{\mathbb V}}
\def\X{{\vec{X}}}
\def\x{{\vec{x}}}

\newtheorem{thm}{Theorem}

\newtheorem{Assumption}{Assumption}

\usepackage{xpatch}
\xpatchcmd{\algorithmic}{\setcounter}{\algorithmicfont\setcounter}{}{}
\providecommand{\algorithmicfont}{\small}

\theoremstyle{definition}

\begin{document}


\title{Nonparametric Conditional Density Estimation in a High-Dimensional Regression Setting}
\date{}
\author{Rafael Izbicki\thanks{Department of Statistics, Federal University of São Carlos, Brazil.} \ and Ann B. Lee\thanks{Department of Statistics, Carnegie Mellon University, USA.}}

\maketitle 

\begin{abstract} 
In some applications (e.g., in cosmology and economics), the regression $\E[Z|\vec{x}]$ is not adequate to represent the association between a predictor $\vec{x}$ and a response 
  $Z$  because of  multi-modality and asymmetry of $f(z|\vec{x})$; using the full density instead of a single-point estimate can then lead to less bias in subsequent analysis.  As of now, there are no effective ways of estimating $f(z|\vec{x})$ when $\vec{x}$ represents  high-dimensional, complex data. 
In this paper, we propose a new nonparametric estimator of $f(z|\vec{x})$ that adapts to sparse (low-dimensional) structure 
in $\vec{x}$.   By directly expanding $f(z|\vec{x})$ in the eigenfunctions of a kernel-based operator,  we avoid tensor products in high dimensions as well as ratios of estimated densities.
 Our basis functions are orthogonal with respect to the underlying data distribution, allowing fast implementation
and tuning of parameters. 
We derive rates of convergence and show that the method adapts to the intrinsic dimension of the data. We also demonstrate the effectiveness of the series method on images, spectra, and 
  an application to 
 photometric redshift estimation of galaxies.

\end{abstract}

\emph{Disclaimer:} The final, accepted version of this paper is published in the Journal of Computational
and Graphical Statistics, and may be found at the website \\
{http://www.tandfonline.com/doi/abs/10.1080/10618600.2015.1094393}.

  \section{Introduction}
\label{intro}

A challenging problem in modern statistical inference is how to 
 handle complex, high-dimensional data where the covariates 
 can be entire images, spectra, or trajectories. 
Whereas researchers have proposed methods for estimating 
the regression of a random variable $Z \in \mathbb{R}$ 
given a high-dimensional random vector $\vec{X} \in \mathbb{R}^d$, i.e., the conditional mean $\E[Z|\vec{x}]$, there is little statistical literature on the problem of estimating the {\em full} conditional density $f(z|\vec{x})$
given an $i.i.d.$ sample from $(Z,\vec{X})$ when $\vec{X}$ is in high dimensions. Yet, in many modern applications, there are clear advantages to estimating $f(z|\vec{x})$
rather than only the regression curve. The list is long:
The conditional density function can, for example, be used to construct more accurate predictive intervals for new observations \citep{Fernandez-Soto}.  Estimating $f(z|\vec{x})$ is a simple way of performing
nonparametric quantile regression \citep{Takeuchi} of many quantiles simultaneously.  Moreover,  in forecasting and prediction, e.g., in economics \citep{filipovic2012conditional,GneitingKatzfuss}, the conditional density itself is often a key quantity of interest. 
Finally, there are situations where the regression $\E[Z|\vec{x}]$ is simply not informative enough to create good predictions of $Z$, because of {\em multi-modality}, {\em asymmetry} or {\em heteroscedastic noise} in  $f(z|\vec{x})$.

As a case in point, several recent works in 
cosmology \citep{Wittman,Sheldon} have shown that one can significantly reduce systematic errors in cosmological analyses by using the full probability distribution of photometric redshifts $Z$ (a key quantity that relates the distance of a galaxy to the observer) 
given galaxy colors $\vec{x}$ (i.e., differences of brightness measures made at two wavelengths). 
This in turn improves estimates of the parameters that dictate the structure and evolution of our Universe. Indeed, in a review of the current state of data mining and machine learning in astronomy, \citet{Ball} listed working with probability densities as one of the ``future trends'' of the field. We will return to the problem of photometric redshift estimation in Sec.~\ref{ex-sdss}.


Several nonparametric estimators have been proposed to estimate conditional densities when $\vec{x}$
lies in a \emph{low-dimensional} space.
Many of them are based on first estimating $f(z,\vec{x})$ and $f(\vec{x})$ with for example kernel density estimators \citep{Rosenblatt}, and then combining the estimates according 
to $f(z|\vec{x})=f(z,\vec{x})/f(\vec{x})$. 
Very few works, however, attempt to estimate $f(z|\vec{x})$ when $\vec{x}$ has  more than
$d=3$ dimensions. Most methods rely on a dimension reduction of $\vec{x}$ prior to implementation
(e.g., \citealp{Fan2}). As is the case with any data reduction, such a
step can result in significant loss of information. 

In a different attempt to 
reduce the number of covariates, 
\citet{Hall2} propose a method for tuning parameters in kernel density 
estimators that automatically determines which components of $\vec{x}$ are 
relevant to $f(z|\vec{x})$. The method produces good results  
but
 because the method selects a different bandwidth for each covariate, 
  the computational cost becomes prohibitive even for moderate sizes of $n$ and $d$. 
A second framework for reducing the number of covariates has been developed by
\citet{Efromovich3}. He proposes an orthogonal series estimator that automatically performs dimension reduction on $\vec{x}$
when several components of this vector are conditionally independent of the response.
The estimator expands the conditional density as a sum of projections on all possible subspaces of reduced dimension, and it uses shrinkage procedures to estimate each projection. The results are comparable to those from \citet{Hall2}.  Unfortunately, Efromovich's method involves computing $d$ tensor products, and like Hall et al., the tensor approach becomes computationally intractable even for as few as 10 covariates. 
Thus, although high-dimensional inference is an active field, there are still no effective methods for estimating full conditional densities in high dimensions.


 The goal of this paper is to answer the following questions:
 (i) Can one find a nonparametric conditional density estimator that performs well {\em in dimensions of the order of hundreds, or even thousands of variables}? 
 In particular, we will consider naturally occurring data where the dimension $d$ of the data is large but the data often have {\em sparse} structure. ``Sparse'' here refers to a general setting where the 
 underlying distribution $P(\vec{x})$
places most of its mass on a subset $\mathcal{X}$ of $\mathbb{R}^d$ of
small Lebesgue measure. This scenario includes, but is not limited to, hyperplanes, Riemannian submanifolds of $\mathbb{R}^d$, 
and high-density clusters separated by low-density regions. 
 (ii) Would the estimator automatically  adapt to the {\em intrinsic} dimensionality of data with ``sparse structure''? For example, if the data $\vec{x}$ lie on a submanifold in  $\mathbb{R}^d$  with dimension $p \ll d$, the convergence rate of the estimator should depend on $p$ rather than $d$. 

 Here we propose a fully nonparametric estimator that  addresses the issues above. 
  The estimator expands  the conditional density $f(z|\vec{x})$ in terms of the estimated eigenfunctions of a kernel-based operator (Eq.~(\ref{eq::operator})); the eigenfunctions are computed using a data-based  Gram matrix  (Eq.~\ref{gramMatrix}). 
 Our approach has some similarities to \cite{Girolami} who uses Kernel PCA~\citep{Schlokopf}  and series expansions, albeit for {\em unconditional} density estimation and without adapting to sparse structure; \cite{Fu} who use kernel-based mappings for conditional density estimation in a {\em parametric} framework; and \citet{izbickihigh}
 who use the estimated eigenfunctions for density ratio estimation.

\begin{figure}[H]
  \centering
\subfloat{  \includegraphics[page=5,scale=0.194]{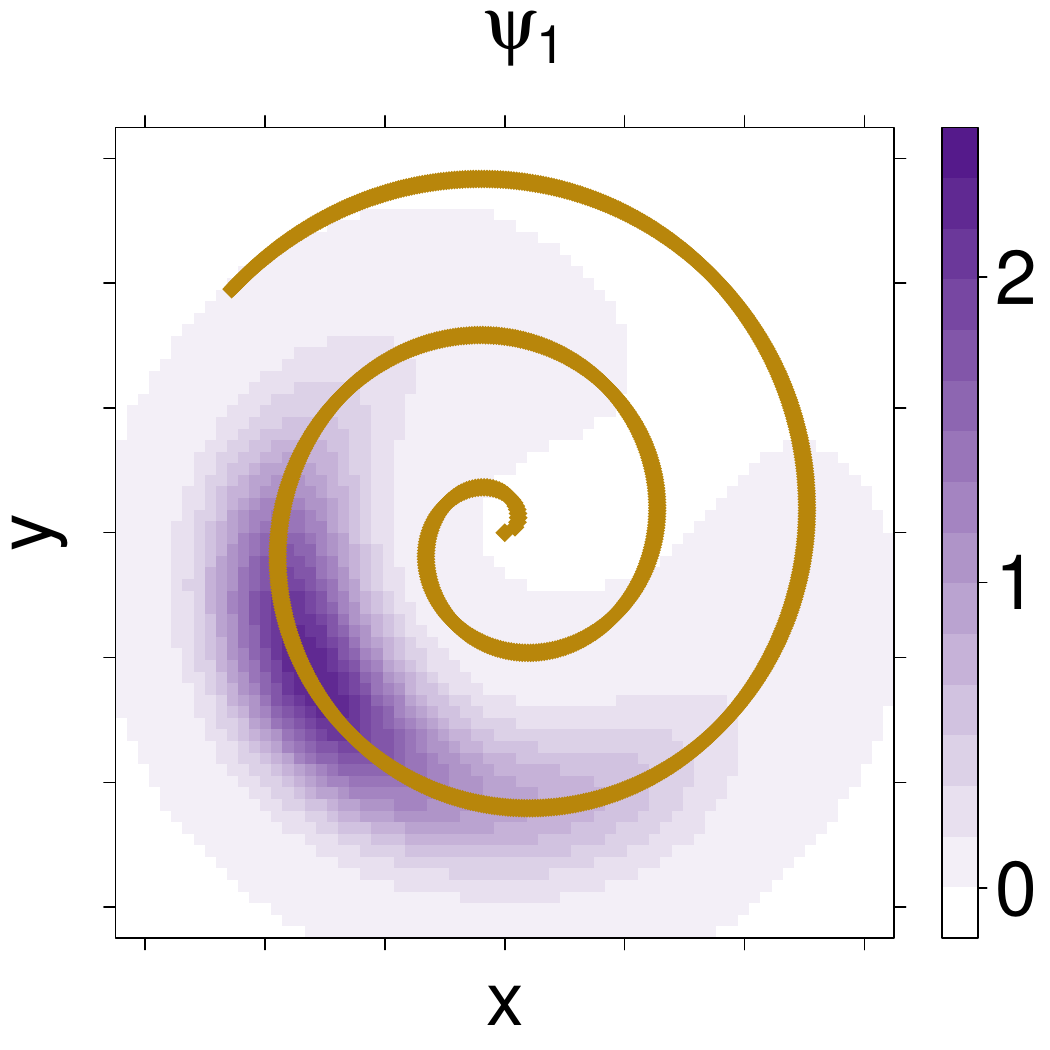}} \hspace{-2.32mm}
\subfloat{  \includegraphics[page=6,scale=0.194]{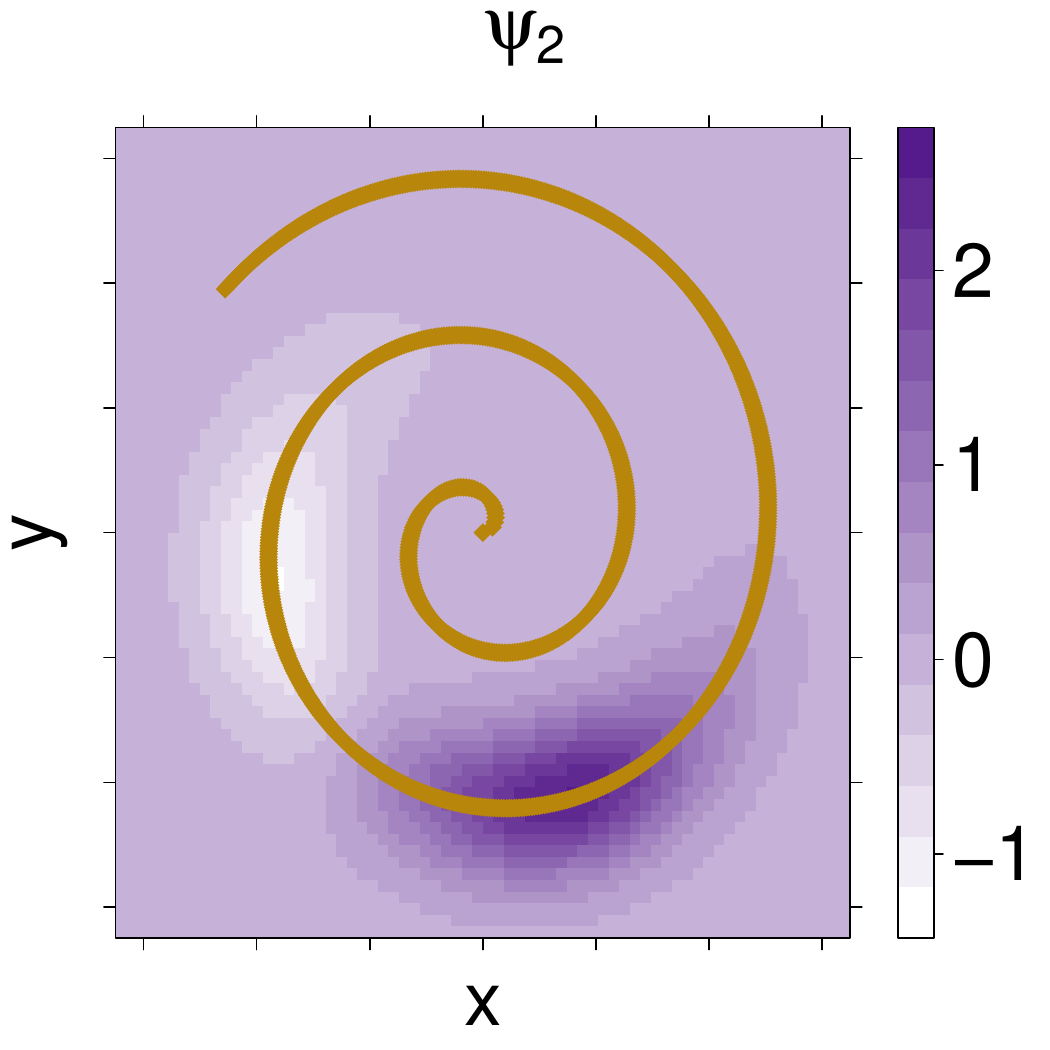}} \hspace{-2.32mm}
\subfloat{  \includegraphics[page=7,scale=0.194]{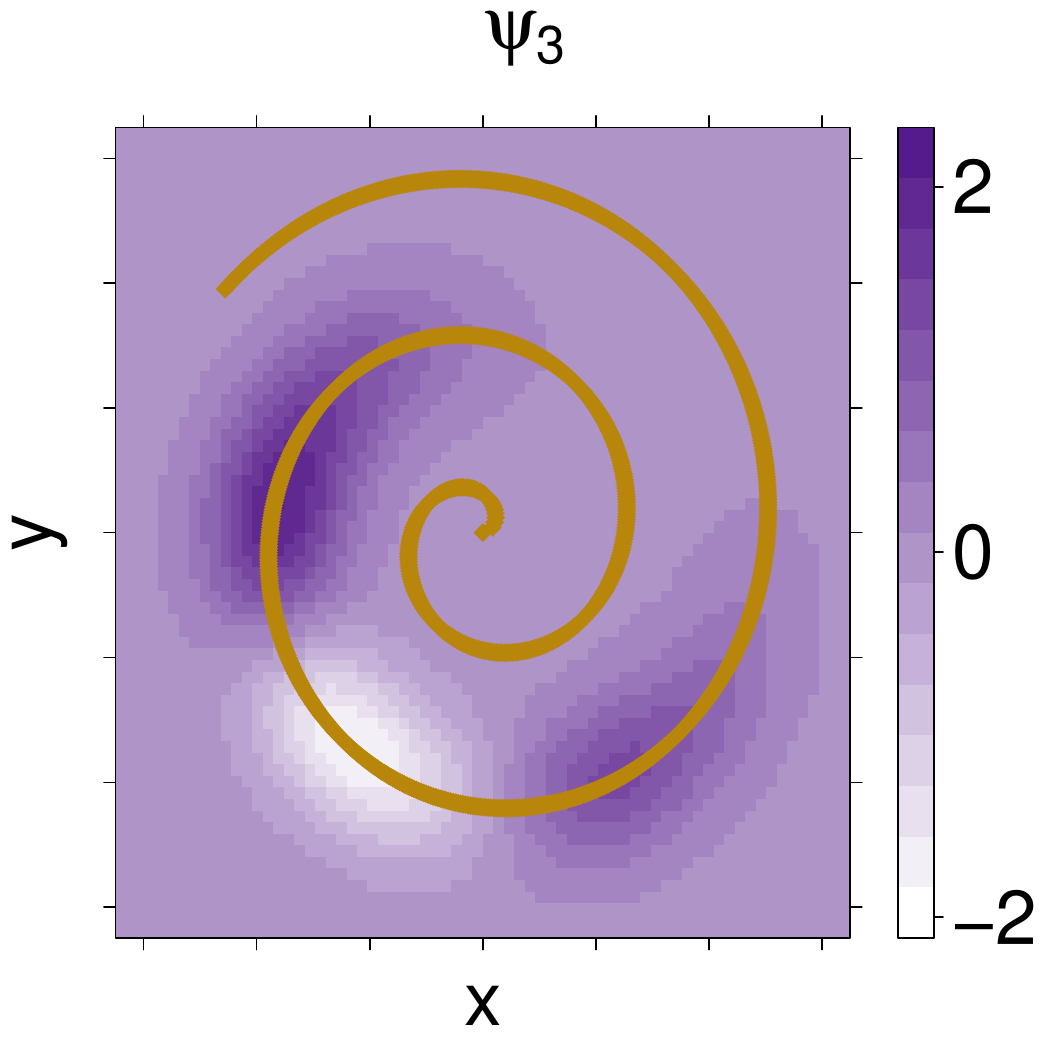}} \hspace{-2.32mm}
\subfloat{  \includegraphics[page=8,scale=0.194]{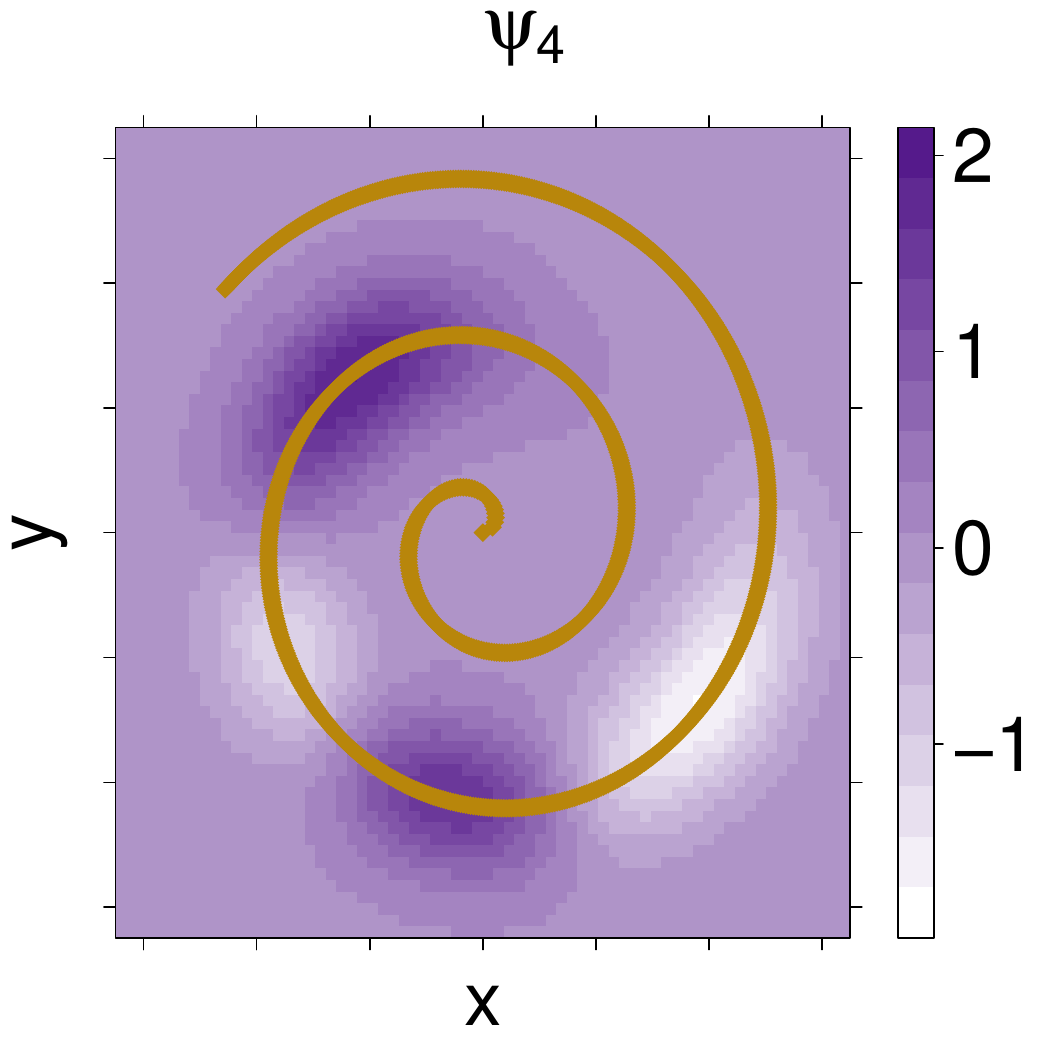}} \hspace{-2.32mm} \\ 
\vspace{-2mm}
  \caption{\footnotesize Level sets of the top eigenfunctions of the Gaussian kernel operator when the domain of the data $\vec{x}=(x,y)$ is on a spiral. The eigenfunctions
  form a Fourier-like basis concentrated around the submanifold,  
  and they are well-suited for approximating smooth functions of $\vec{x}$ on this domain.}
  \label{fig::eigenSpiral}
\end{figure}

It is widely known that, due to the ``curse of dimensionality'' \citep{bellman1961adaptive},
fully nonparametric inference
is difficult in high dimensions without unrealistic amounts of data and computing power. There are several reasons why our series approach still can be effective in high dimensions: 
 (i)  Our computed basis functions  are adapted to the {\em intrinsic}  geometry of the data.
 For example,
 when the domain of the data is close to a submanifold $\Omega \in \mathbb{R}^d$, where $d$ can be large, the eigenfunctions form a Fourier-like basis concentrated around the submanifold
 with lower-order terms smoother than higher-order terms. 
 Fig.~\ref{fig::eigenSpiral} shows an example.
   If $f(z|\vec{x})$  is smooth relative to this domain, then we only need a few eigenfunctions to approximate the unknown density.  
As we shall see in Sec.~\ref{sec-theory}, this yields convergence rates that depend on the intrinsic rather than the ambient dimension of the data.

(ii) 
Our basis functions are {\em orthogonal with respect to $P(\x)$}, the underlying data distribution, instead of orthogonal with respect to the Lebesgue measure of the ambient space as in traditional  orthogonal series methods. Because of this property, we can quickly estimate the expansion coefficients in the conditional density estimator by taking empirical averages (Eq.~(\ref{beta})). The tuning of parameters is fast. We do not need cumbersome tensor products in high dimensions, nor do we need to recompute the expansion coefficients when varying the number of terms in the series. (iii) Finally, our proposed method {\em directly} estimates $f(z|\vec{x})$ 
 and avoids dividing two estimated densities as in  $\widehat{f}(z,\vec{x})/\widehat{f}(\vec{x})$. The latter two-step approach is common in other approaches but can magnify estimation errors and lead to poor estimates \citep{ chagny2013warped}, especially in high dimensions. 
 Estimating $f(\vec{x})$  can, in fact, be harder than estimating $f(z|\vec{x})$ when $f(\vec{x})$ is less 
smooth than $f(z|\vec{x})$; see \citet{Efromovich3}.


So far, orthogonal series methods have been limited to settings with only a few covariates. Here we present theoretical and empirical evidence that series methods 
 can indeed be effective in dimensions with upwards of $10^3$ variables with the right choice of basis. This work opens up a whole range of possibilities for using Fourier methods and orthogonal series for {\em estimating functions} on complex non-standard data in high dimensions. As a by-product of our spectral approach, we also have a natural means for {\em visualizing}  and {\em organizing} such data.
 Figure 1 in the appendix  shows an embedding of astronomy data into a lower-dimensional space, where the first few basis functions are used as coordinates.

  Sec.~\ref{cders} describes the spectral series method. 
  Sec.~\ref{sec-theory} gives theoretical guarantees on our estimator. In Sec.~\ref{sec-application}, we compare the  performance of spectral series with other estimators for a wide range of simulated and real-world  data. 
We conclude in Sec.~\ref{sec-conclusions}. 


\section{Methodology}
\label{cders}

In this paper, we propose a new nonparametric conditional density estimator that performs well in high dimensions and that automatically adapts to the intrinsic dimension of the data. The main idea is to project the conditional density $f(z|\x)$ onto the data-dependent eigenfunctions of a kernel-based operator. We then take advantage of the orthogonality of the basis for fast computation and tuning of parameters. The details are as follows: 

 Let $(Z_1,\vec{X}_1),\ldots,(Z_n,\vec{X}_n)$ denote an i.i.d. \hspace{-3mm} sample, where $\vec{X}_i \! \in \! \mathcal{X} \! \subseteq \! \mathbb{R}^d$,
and the domain of $z$ is bounded; for simplicity, we assume $Z_i \! \in \!  [0,1]$.
Let $P(\vec{x})$ be the distribution of $\vec{X}_i$.  

\noindent {\bf Projecting $f(z|\vec{x})$ onto a spectral basis.} Let $K(\vec{x},\vec{y})$ be a Mercer kernel; that is, $K$ is bounded, symmetric,  
and positive definite. $K$ measures the similarity between pairs of data points.
A popular choice in kernel machine learning is the Gaussian kernel, 
$K(\vec{x},\vec{y}) =\exp{\left( \frac{-d^2(\vec{x},\vec{y})}{ 4\epsilon} \right)},$ 
where $d(\cdot,\cdot)$ is the Euclidean
distance in $\mathbb{R}^d$ and $\epsilon$ is a bandwidth chosen
according to Sec.~\ref{lossTuning}.
As in spectral clustering (e.g, \citealp{Shi}), we define an integral operator $\textbf{K} \! : \! \mathfrak{L}^2(\mathcal{X},P) \longrightarrow \mathfrak{L}^2(\mathcal{X},P)$
by
\begin{align}
\label{eq::operator}
\textbf{K}(h)(\vec{x})=\int_\mathcal{X} \! K(\vec{x},\vec{y})h(\vec{y})dP(\vec{y}).
\end{align}
The operator $\textbf{K}$ 
has a countable number of eigenfunctions $\psi_1,\psi_2,\ldots$
with respective eigenvalues $\lambda_1 \geq \lambda_2 \geq \ldots \geq 0$ \citep{Mink}. 
%
These eigenfunctions form an {\em adaptive} orthonormal basis of $\mathfrak{L}^2(\mathcal{X},P)$ --
the Hilbert space of square integrable functions with domain $\mathcal{X}$ 
and norm $||g||^2_{P}=\langle g,g\rangle_{P}=\int_\mathcal{X} |g(\vec{x})|^2dP(\vec{x})$ \citep{Minh2}.  More precisely, the eigenfunctions are  orthonormal with respect to the  data distribution $P(\vec{x})$,   
$$\int_\mathcal{X} \! \psi_i(\vec{x})\psi_j(\vec{x})dP(\vec{x})=\delta_{i,j}\overset{\mbox{\tiny{def}}}{=}\I(i=j), $$ and they can be used to approximate smooth functions of $\vec{x}$. 


 The central idea of our spectral series estimator is to project $f(z|\vec{x})$, which is a function of both  $\vec{x}$ and $z$, onto only {\em one} tensor product 
  \begin{equation}\Psi_{i,j}(z,\vec{x})=\phi_i(z)\psi_j(\vec{x}),\ i,j\in \mathbb{N}, \label{eq:tensorproduct_basis}\end{equation}
where $\{\psi_j\}_{j \in \mathbb{N}}$ is the spectral basis on $\mathcal{X}$, and 
$\{\phi_i\}_{i\in \mathbb{N}}$ is a suitable orthonormal basis on the domain of $z$, 
 $\int_{[0,1]} \phi_i(z)\phi_j(z)dz=\delta_{i,j}$. Because $z$ is scalar, 
there is a wide range of possibilities (see Remarks 1). In this paper, we use the standard Fourier basis.
 On the other hand, classical series estimators~\citep{efromovich1999nonparametric}, as well as the recent conditional density estimator by  \cite{Efromovich3}, involve as many as $d$ tensor products of functions in $\Re$, making them computationally intractable even for $d=10$ covariates. 

By projecting onto the spectral tensor product basis, we have the series expansion
\begin{eqnarray}
\label{eq-proj}
f(z|\vec{x})=\sum_{i,j}\beta_{i,j}\Psi_{i,j}(z,\vec{x}) \,,
\end{eqnarray}
where the coefficients $\beta_{i,j}$ take a particularly simple form:
 Because $\psi$ is orthogonal with respect to the data 
distribution, and because $\phi$ is orthogonal with respect to 
Lebesgue measure, the coefficients are simply expectations over the joint 
distribution of $\vec{X}$ and $Z$, 
$$\beta_{i,j}=\iint f(z|\vec{x})\Psi_{i,j}(z,x)\:dP(\vec{x})dz=\iint \Psi_{i,j}(z,\vec{x})\:dP(z,\vec{x})=\E[\Psi_{i,j}(Z,\vec{X})].$$ 


\noindent {\bf Computing the conditional density estimator from data.} As $P(\vec{x})$ is unknown, we need to estimate the $\psi_j$'s. We compute the eigenvectors of the Gram matrix
\begin{equation}
 \label{gramMatrix}
\left[ K_\x\left(\vec{x}_i,\vec{x}_j\right) \right]_{i,j=1}^{n}.  
 \end{equation}
Let 
$\widetilde{\psi}_j:=\left(\widetilde{\psi}_j(\vec{x}_1),\ldots,\widetilde{\psi}_j(\vec{x}_{n})\right)$
be the $j$-th eigenvector of the matrix in Eq.~\ref{gramMatrix}, and let $\widehat{l}_j$ be its associated eigenvalue. 
 We sort the eigenvectors by decreasing order of eigenvalues, and normalize them so that $\sum_{k=1}^n\widetilde{\psi}^2_j(\vec{x}_k)=1$. 
 One can show that the Nystr\"om extension \citep{Drineas05onthe}
 $\widehat{\psi}_j(\vec{x})=\frac{\sqrt{n}}{\widehat{l}_j}\sum_{k=1}^n \widetilde{\psi}_j(\vec{x}_k) K(\vec{x},\vec{x}_k)$ is a consistent estimate of $\psi_j$ \citep{Bengio}. 

We define the spectral series estimator 
\begin{align}
\label{eq-finalEstimator}
\widehat{f}(z|\vec{x})=\sum_{i=1}^I\sum_{j=1}^J\widehat{\beta}_{i,j}\widehat{\Psi}_{i,j}(z,\vec{x}),
\end{align}
where the parameters $I$ and $J$ 
 control the bias/variance tradeoff, 
 \begin{align}
\label{PsiEstimate}
\widehat{\Psi}_{i,j}(z,\vec{x})= \phi_i(z) \widehat{\psi}_j(\vec{x})
\end{align}
 is the estimate of $\Psi_{i,j}(z, \vec{x})$, and $\widehat{\beta}_{i,j}$ are empirical averages,
\begin{align}
\label{beta}
 \widehat{\beta}_{i,j}=\frac{1}{n}\sum_{k=1}^n \widehat{\Psi}_{i,j}(z_k,\vec{x}_k).
\end{align}
Because of the orthogonality property of the basis, it is fast to cross-validate over $I$ and $J$. There is essentially no need to update the coefficients $\widehat{\beta}_{i,j}$ when varying $I$ and $J$. We refer to Sec.~\ref{lossTuning} for details on tuning the parameters of the estimator.

\comment{Next we estimate the expansion coefficients  by empirical averages:
\begin{align}
\label{beta}
 \widehat{\beta}_{i,j}=\frac{1}{n}\sum_{k=1}^n \widehat{\Psi}_{i,j}(z_k,\vec{x}_k),
\end{align}
where 
\begin{align}
\label{PsiEstimate}
\widehat{\Psi}_{i,j}(z_k,\vec{x}_k)=\phi_i(z_k)\widehat{\psi}_j(\vec{x}_k) 
\end{align}
is the estimate of $\Psi_{i,j}(z_k,\vec{x}_k)$.
Finally, we define our 
spectral series estimator of $f(z|\vec{x})$ according to 
\begin{align}
\label{eq-finalEstimator}
\widehat{f}(z|\vec{x})=\sum_{i=1}^I\sum_{j=1}^J\widehat{\beta}_{i,j}\widehat{\Psi}_{i,j}(z,\vec{x}).
\end{align}
The parameters $I$ and $J$ 
 control the bias/variance tradeoff: By decreasing their values, we decrease the variance,
but increase the bias of the estimator.  In Sec.~\ref{lossTuning}, we explain how to choose the tuning parameters. \\ 
}

\comment{
Notice that to use the spectral series method we do not need $d$ tensor as in \citet{efromovich1999nonparametric}. Instead, we only need
\emph{one} tensor product, the product between the basis for $\x$ and the basis for $z$. 
This is because the basis $\psi(\x)$ used by Efromovich is built using $d$
tensor products of functions in $\Re$, whereas here we use the spectral decomposition of the Gram matrix.
This is what allows the spectral series estimator to be used even when $d$ is large with no computational difficulties.}
\vspace{2mm}

 {\small 
  \noindent \emph{Remarks -- further extensions of the spectral series method:} 
\begin{enumerate}[leftmargin=*]
 

\item  Spectral series are more flexible than kernel smoothers because
 one can model the density $f(z|\vec{x})$
as a function of $z$ using a variety of different bases~\citep{efromovich1999nonparametric}; for example, 
  Fourier bases or, in the case of 
  spatially inhomogeneous densities in $z$, wavelet bases. 
In Sec.~\ref{ex-zip}, where the response $Z$ takes values on a {\em discrete} set $\{1,\ldots,p\}$, we introduce the indicator basis $\phi_i(z)=\I(z=i),\ i=1,\ldots,p$  with inner product
 $\langle f,g\rangle=\sum_{i=1}^p f(i)g(i)$.

\item  
By choosing an appropriate kernel (or data similarity matrix),  
 spectral series can handle {\em different types 
of covariate data} $\x$; e.g., 
 SNP genetic data~\citep{lee2010spectral},
 functional data, circular data, and abstract objects on a graph. 
 \citet{scholkopf2001learning} list other kernels and some of their advantages and disadvantages. 
Note that given a set of reasonable candidate kernels, one can choose ``the best kernel'' with the smallest estimated loss according to Eq.~(\ref{lossEmpirical}).

 \item The spectral series framework naturally extends to  {\em semi-supervised learning} (SSL)
  ~\citep{zhu2009introduction} where besides the labeled sample $(\X_1,Z_1),\ldots,(\X_n,Z_n)$ there are
  additional unlabeled data; i.e., data  $\X_{n+1},\ldots,\X_{n+m}$ where the covariates $\vec{x}$ but not the labels $z$ are known.
    By including the unlabeled data in the Gram matrix (Eq.~\ref{gramMatrix}), one can better estimate the eigenfunctions $\psi_j$ and, hence, the conditional density $f(z|\x)$; see Sec.~\ref{sec-theory} for theory.

\item   In the spectral clustering literature~\citep{von2007tutorial}, there exist several normalized variants of the operator in Eq.~(\ref{eq::operator}). To simplify our proofs, we will use the normalized \emph{diffusion operator} \citep{Lee} defined in Appendix A.3.
As shown in Sec.~\ref{sec-application}, the empirical performance for spectral series CDE is similar for the normalized and unnormalized variants of the kernel operator. 

\end{enumerate}
}

\subsection{Loss Function and Tuning of Parameters}
\label{lossTuning}

For a given estimator $\widehat{f}(z|\vec{x})$, we measure the discrepancy between  $\widehat{f}(z|\vec{x})$ and $f(z|\vec{x})$ via the loss function 
\begin{align}
\label{loss} \nonumber 
 L(\widehat{f},f) &= \iint \left(\widehat{f}(z|\vec{x})-f(z|\vec{x})\right)^2dP(\vec{x})dz \\ 
 &=\iint \widehat{f}^2(z|\vec{x})dP(\vec{x})dz-2\iint \widehat{f}(z|\vec{x})f(z,\vec{x}) d\vec{x}dz+C, 
\end{align}
where $C$ is a constant that does not depend on the estimator. 
\comment{This measure is appropriate for high-dimensional problems where data are sparse: }
The weighting by $P$ reflects the fact that we are primarily interested in accurately estimating the density at $\vec{x}$'s that occur frequently.


To tune parameters, we split the data into a training and a validation set.  
For each configuration of the tuning parameters ($I$, $J$ and $\epsilon$) on a grid, we use the training set to estimate the coefficients $\beta_{i,j}$ according to Eq.~(\ref{beta}). We then use the validation set $(z'_1,\vec{x}'_1),\ldots, (z'_{n'},\vec{x}'_{n'})$ to 
estimate the loss (\ref{loss}) (up to the constant $C$) according to:
\begin{align}
\label{lossEmpirical}
\widehat{L}(\widehat{f},f)=\sum_{i=1}^I\sum_{j=1}^J\sum_{m=1}^J\widehat{\beta}_{i,j}\widehat{\beta}_{i,m}\widehat{W}_{j,m}-2\frac{1}{n'}\sum_{k=1}^{n'} \widehat{f}(z'_k|\vec{x}'_k),
\end{align}
where 
$\widehat{W}_{j,m}=(n')^{-1}\sum_{k=1}^{n'}\widehat{\psi}_j(\vec{x}'_k)\widehat{\psi}_m(\vec{x}'_k).$
We choose the tuning parameters 
with the smallest estimated loss $\widehat{L}(\widehat{f},f)$.

 Algorithm 1 summarizes our procedure. Naturally, if the sample size is small, one
can use cross-validation \citep{Hastie:EtAl:09} instead of data splitting. 
As mentioned, the estimated coefficients $\widehat \beta_{i,j}$ do not depend on $I$ and $J$. It follows that if we compute  $\widehat \beta_{i,j}$ for all $i \leq I_{\rm max}$ and $j \leq J_{\rm max}$ (where $I_{\rm max}$ and $J_{\rm max}$ are the largest values of $I$ and $J$ on the grid),  
  then we do not need to recompute these coefficients for other configurations of $I$ and $J$.
 This gives spectral series a clear competitive edge in terms of speed relative
  least squares procedures, such as,  
\cite{kanamori2012statistical}.

  \begin{algorithm}
 \caption{ \small Tuning the Spectral Series Conditional Density Estimator}\label{algorithmCondReg}
 \algorithmicrequire \ {\small Training data; validation data; grid over $\epsilon$, $I$ and $J$. } 
 
 \algorithmicensure \ {\small Estimator $\widehat{f}(z|\vec{x})$}
 \begin{algorithmic}[1]
    \ForAll{$\epsilon$}
       \State calculate the eigenvectors $\widetilde{\psi}=\widetilde{\psi}_{\epsilon}$ of the Gram matrix \Comment{Eq.~(\ref{gramMatrix})}
       \State estimate the eigenbasis $\widehat{\Psi}_{i,j}$ \Comment{Eq.~(\ref{PsiEstimate})}
       \State estimate the coefficients $\widehat{\beta}_{i,j}$ \Comment{Eq.~(\ref{beta})}
       \ForAll{$I,J$}
 	\State calculate the estimated loss $\widehat{L}(\widehat{f}_{\epsilon,I,J},f)$ \Comment{Eq.~(\ref{lossEmpirical})} 
       \EndFor
    \EndFor
     \State Define $\widehat{f}=\arg \min_{\widehat{f}_{\epsilon,I,J}(z|\vec{x})} \widehat{L}(\widehat{f}_{\epsilon,I,J},f)$
    \State \textbf{return} $\widehat{f}(z|\vec{x})$
 
 \end{algorithmic}
 \end{algorithm}
\subsection{Normalization and Spurious Bumps}
\label{normalization}

In the statistics literature, there are many approaches for transforming a general density estimate $\widehat{f}$
into a bona fide density $\widetilde{f}$ that is non-negative and that integrates to one.   For an overview and theoretical guarantees, 
we refer the reader to \citealt{hall1993correcting,efromovich1999nonparametric,Glad,wasserman}.
We found that the following procedure gave good results for our data: Let 
$\widehat{f}_{\max}(z|\vec{x})=\max\left\{0,\widehat{f}(z|\vec{x})\right\}.$
 If $\int \! \widehat{f}_{\max}(z|\vec{x})dz\geq 1$, then for each $\vec{x}$ and $z$, define  $\widetilde{f}(z|\vec{x})=\max\{0,\widehat{f}(z|\vec{x})-\xi\},$
where $\xi$ is such that 
$\int \! \widetilde{f}(z|\vec{x}) dz = 1.$ 
 If $\int \! \widehat{f}_{\max}(z|\vec{x})dz< 1$, then define 
 $\widetilde{f}(z|\vec{x})=\widehat{f}_{\max}(z|\vec{x})/ \int \! \widehat{f}_{\max}(z|\vec{x}) dz.$
Following \citet{efromovich1999nonparametric}, we 
  also remove a bump in the interval $[a,b]$ when 
$\int_a^b \widetilde{f}(z|\vec{x})dz<\delta$,  as 
 small spurious bumps can arise if one approximates the flat parts of the underlying density with a finite series of oscillating functions. 
We treat $\delta$ as a tuning parameter, and choose the optimal value $\delta^*$
that minimizes the estimated loss in Eq.~(\ref{lossEmpirical}).
 To speed up the computations, we take on a greedy approach and tune $\delta$ after determining the other tuning parameters.

\subsection{Scalability}
\label{sec::improv}

 The spectral series estimator, even in its naive implementation, is faster than most traditional approaches, especially in high dimensions.  The only computation that depends on the dimension $d$ is the construction of the Gram (similarity) matrix. Once this matrix has been constructed, the eigendecomposition takes the same amount of time for all values of $d$. 
 Nevertheless, simple improvements can 
further reduce the complexity of the spectral series method.  
By using Randomized SVD \citep{Halko}, 
one can speed up the eigendecomposition of the Gram matrix, $\boldsymbol{G}$, from $\BigO{n^3}$ to roughly
 $\BigO{n^2}$, when $J \ll n$, with little decrease in statistical performance. 
 In addition, one can reduce the memory complexity of spectral series by making $\boldsymbol{G}$ sparse.  For local kernels (e.g.,
 the Gaussian kernel), 
 the matrix $\boldsymbol{G}$ can be stored with less memory after a simple thresholding; i.e, after setting all entries with $K(\vec{x}_i,\vec{x}_j)$ less than a small user-specified value $\xi>0$ to $0$.
The parameter $\xi$
controls the trade-off between evaluation precision and memory complexity.
This is illustrated in Sec.~\ref{ex-sdss}, where 
we will revisit the topic of scalability with numerical examples of photometric redshift estimation. 
 Further improvements, not explored in this work, include SVD with multi-processor architectures \citep{Halko}, fast nearest neighborhood computations, such as, randomized partition trees \citep{dasgupta2013randomized}, cover trees \citep{beygelzimer2006cover}, approximate NN methods (e.g., \citealp{nolen2013approximate}), and 
 (parallelizable) multi-trees \citep{gray2000n,boyer2007parallel} that trade off evaluation precision and computational speed.

\section{Theory}
\label{sec-theory}

Next we provide theoretical guarantees that the estimator $\widehat{f}(z|\vec{x})$ is not too far from the true density $f(z|\vec{x})$; i.e. we compute  bounds on the loss (\ref{loss}) of the estimator in Eq.~(\ref{eq-finalEstimator}). Our assumptions are:

\begin{Assumption}
\label{assumpfinite}
$\int f^2(z|\vec{x})dP(\vec{x})dz < \infty$.
\end{Assumption}

\begin{Assumption}
\label{assumpPhi}
 $M_{\phi}\overset{\mbox{\tiny{def}}}{=}\sup_z \sup_i |\phi_i(z)| < \infty$.
\end{Assumption}

\begin{Assumption} 
 \label{decayEigenvalues}
$\lambda_1>\lambda_2>\ldots>\lambda_J>0$.
\end{Assumption}



\noindent Assumption \ref{assumpfinite} implies that it is possible to expand $f$ in
 the basis $\Psi$. 
Assumption \ref{assumpPhi} depends on the choice of basis for $z$; it holds, e.g., for cosine or Fourier bases.
Assumption \ref{decayEigenvalues} allows uniquely defined eigenfunctions; see, e.g., \cite{Zwald} on how to proceed if the eigenvalues are degenerate. 

To estimate $f(z|x)$, we need $f$ to belong to a set of functions which are not too ``wiggly''. 
For every $ s>\frac{1}{2}$ and $0<c<\infty $, let $W_{\phi}(s,c)$ denote the Sobolev space $W_{\phi}(s,c)=\{f \!= \!\sum_{i\geq 1} \theta_i \phi_i \!:  \! \sum_{i\geq 1} a_i^2 \theta^2_i \leq c^2 \}$, where $a_i \! \sim \! (\pi i)^s$. For the Fourier basis
 $\phi$, this is the standard definition of Sobolev space \citep{wasserman}; it is  the
 space of functions that have their $s$-th weak derivative bounded by $c^2$ and integrable in $L^2$.
We enforce smoothness in the $z$-direction by requiring $f(z|\x)$ to be in a Sobolev space for all $\vec{x}$,
\begin{Assumption} [Smoothness in $z$ direction]
 \label{assump-sobolevZ} $\forall \vec{x} \! \in \! \mathcal{X}$, 
$f(z|\x) \! \in \! W_{\phi}(s_\vec{x},c_\vec{x}),$
 where $f(z|\x)$ is viewed as a function of $z$, and $s_\vec{x}$ and $c_\vec{x}$ are such that 
$\inf_\vec{x} s_\vec{x}\overset{\mbox{\tiny{def}}}{=}\beta>\frac{1}{2}$ and $\int_\mathcal{X} c_\vec{x}^2dP(\vec{x})<\infty$.
\end{Assumption}
\noindent The quantities $\beta$ and $\int_\mathcal{X} c_\vec{x}^2dP(\vec{x})$ are used to 
link the parameters $s_\vec{x}$ and $c_\vec{x}$ that control the degrees of smoothness at different values of $\vec{x}$. Larger values of $\beta$ indicate smoother functions. 

We also assume that $f(z|\vec{x})$ is smooth in the $\vec{x}$ direction. 
 We measure smoothness via a {\em density-weighted} operator: 
 Let
$
S(A) =\Lim{\epsilon \longrightarrow 0} \frac{\int_A p_\epsilon(\x) dP(\x)}{\int p_\epsilon(\x) dP(\x)}$
be a smoothed version of $P$ \citep{Lee}. 
 We assume:  

\begin{Assumption}
 \label{assump-AssumptionPrime}
 {\normalfont(Smoothness in $\vec{x}$ direction)}  $\forall z \in [0,1]$ fixed,
 $\int_{\mathcal{X}}  \|\nabla f(z|\x) \|^2 dS(\x)<c_z,$
 where $c_z$ is such that $\int_{[0,1]} c_z dz  < \infty.$ 
\end{Assumption}

\noindent This 
 measure of smoothness can be seen as a generalization of Sobolev differentiability to sparse structures in high dimensions. 
In Appendix A.5 we prove:

\begin{thm}
\label{thm-boundManifold} 
 Let $\widehat{f}_{I,J}(z|\vec{x})$ be the spectral series estimator from Sec.~\ref{cders} with cutoffs $I$ and $J$ and the eigenfunctions
 of the normalized operator of Appendix A.3
  as a basis. Assume \ref{assumpfinite}-\ref{assump-AssumptionPrime}.
Suppose that 
the kernel $K=K_\epsilon^{*}$  is 
renormalized according to 
$K_\epsilon^{*}(\x,\vec{y})=\frac{K_\epsilon(\x,\vec{y})}{p_\epsilon(\x)p_\epsilon(\mathbf{y})}$.
Then, 
if the support of the data is on a manifold with intrinsic
dimension $p$, 
under the regularity conditions in the appendix, we have that,  for width $\epsilon \asymp n^{-2/(p+4)}$, 
$$L(\widehat{f}_{I,J}, f)  =   O \left(\frac{1}{J^{2/p}}\right) + O\left(\frac{1}{I^{2\beta}}\right)  + IJ^{2\left(1-\frac{1}{p}\right)} O_P\left(\frac{\log n}{n}\right)^{\frac{2}{p+4}}.$$
It is then optimal to choose $I \asymp  n^{\frac{4}{p(p+4)(2/p+4\beta)}} $ and $J \asymp n^{\frac{4\beta}{(p+4)(2/p+4\beta)}}$, in which case the upper bound becomes 
$$O_P\left(n^{\frac{-4\beta}{(p+4)(1+2\beta p)}}\right)=O_P\left(n^{-\frac{1}{O\left(p^2\right)}}\right).$$
In a SSL learning setting
with additional unlabeled data $m \rightarrow \infty$ (see Remark 3 in Sec.~\ref{cders}), the loss reduces to
$$L(\widehat{f}_{I,J}, f)  =   O \left(\frac{1}{J^{2/p}}\right) + O\left(\frac{1}{I^{2\beta}}\right)  + IJO_P\left(\frac{1}{n}\right),$$
in which case it is optimal to choose $I \asymp n^{\frac{1}{2\beta+1+p\beta}}$ and $J \asymp n^{\frac{p\beta}{2\beta+1+p\beta}}$. This yields the  rate  
$$O_P\left(n^{-\frac{2\beta}{2\beta+1+\beta p}}\right)=O_P\left(n^{-\frac{1}{O\left(p\right)}}\right).$$
\end{thm}

Theorem \ref{thm-boundManifold} shows that the rate of convergence of the spectral series estimator depends only on the {\em  intrinsic}
dimension $p$, which can be much smaller than the ambient dimension $d$.  In the limit of infinite unlabeled data, our rate is of 
the form $O_P\left(n^{-1/O\left(p\right)}\right)$. Compare this result to the standard rates for nonparametric conditional density estimators which are of the form $O_P\left(n^{-1/O\left(d\right)}\right)$ \citep{Hall2}.
In particular, in the isotropic setting (where $\beta=1$ due to Assumption \ref{assump-AssumptionPrime}), the series estimator achieves the minimax rate
$O_P\left(n^{-2/(2+(1+p))}\right)$ for estimators in $p$+1 dimensions.
On the other hand, if there is no unlabeled data, we guarantee
$O_P\left(n^{-1/O\left(p^2\right)}\right)$ rates. This bound may be overly pessimistic as it assumes that the eigenvectors need to be accurately estimated. Indeed, our empirical experiments indicate that spectral series (with approximate eigenvectors) perform better or as well as the nearest neighbor method which is minimax optimal in regression \citep{kpotufe2011k}.    Notice, however, that when $p \ll d$, this is {\em still} considerably better than $O_P\left(n^{-1/O\left(d\right)}\right)$.

Note that  spectral series use a different mechanism to overcome the curse-of-dimensionality compared to the estimators from \citet{Hall2} and \citet{Efromovich3}. The latter estimators perform well when the conditional density $f(z|\x)$ of the response $Z$ depends on a small subset of the original covariates $\X$; indeed, the rates are of the form 
$O_P\left(n^{-1/O\left(r\right)}\right)$, where $r$ is the number of relevant covariates in the density estimation.
Spectral series, on the other hand, achieve better rates of convergence if the {\em intrinsic} dimension of the data distribution $P(\x)$ is smaller than the ambient dimension $d$ (see Theorem \ref{thm-boundManifold}).
We refer to the appendix for additional theory and proofs. 
Main results include Theorem 1 in A.4, which is a bound on spectral series for the standard RKHS setting with a fixed kernel, and Theorem 4 in A.5, which is a bound on the estimator for a kernel with {\em varying} variance.

\comment{
Our estimator is
able to overcome the curse-of-dimensionality in a different sense from those estimators:
while 
the estimators from 
\citet{Hall2} and \citet{Efromovich3} overcome the curse when
the number of relevant covariates is small,
the spectral series estimator achieves better rates
of convergence if the intrinsic dimension
of the data is smaller than its ambient dimension.}

\vspace{-6mm}

\section{Numerical Examples}
\label{sec-application}

Next we investigate how different approaches to CDE perform on 
simulated data as well as images of digits, galaxy spectra, and photometric data from astronomical surveys. 
Except for two estimators (\textit{LS} and \textit{KDE}$_{\mbox{\tiny Tree}}$), we choose the tuning parameters according to Sec.~\ref{lossTuning}. More specifically:
\begin{itemize}[leftmargin=*]
   \item  \textit{Series} and \textit{Series$_{\mbox{\tiny Diff}}$} are spectral series estimators with a radial Gaussian kernel in $\x$ and a Fourier basis in the $z$-direction.  \textit{Series}  is based on the unnormalized kernel operator, whereas \textit{Series$_{\mbox{\tiny Diff}}$} uses the (normalized) diffusion operator 
from Appendix A.3.

  \item  \textit{LS} is the  direct least squares conditional density estimator of \cite{sugiyama2010conditional},
 implemented with the MATLAB code and the cross-validation procedure provided by the authors. 
 Like  
  \textit{Series}, the estimator consists of 
  a direct expansion of $f(z|\vec{x})$ in  functions $\psi$. However, the basis functions in  \textit{LS} are not 
  adapted to the underlying data geometry, nor do they form a Hilbert basis for functions on the data.

\item \textit{KDE} is the kernel density estimator $\widehat{f}(z|\x):=\widehat{f}(z,\x)/\widehat{f}(\x)$, where  $\widehat{f}(z,\x)$ and $\widehat{f}(\x)$ 
are standard multivariate normal kernel density estimators. The kernel bandwidth is the same for all components of $\x$,  which have been rescaled to have the same mean and variance. 

 \item \textit{KDE}$_{\mbox{\tiny Tree}}$ is the kernel density estimator $\widehat{f}(z|\x):=\widehat{f}(z,\x)/\widehat{f}(\x)$,
 where the kernel density estimators $\widehat{f}(z,\x)$ and $\widehat{f}(\x)$ use a {\em different} bandwidth for each component of $\x$, but the  bandwidth vector is the same for 
the numerator and the denominator. 
 We use the R package NP \citep{np} to implement the estimator.   Because the cross-validation procedure in \citealt{Hall2} is computationally intractable for large sample sizes and high dimensions, we instead use the R package implementation with kd-trees and likelihood-cross-validated bandwidths \citep{gray2003nonparametric,Holmes}.

\item  \textit{KNN} is a kernel nearest neighbors approach \citep{Lincheng} to conditional density estimation, defined as
 $\widehat{f}(z|\vec{x})\propto \sum_{k\in \mathcal{N}_N(\vec{x})}   K_{\epsilon}\left(z-z_k\right)$,
 where $\mathcal{N}_N(\vec{x})$ is the set of the $N$ closest neighbors to $\x$ in the training set, and
$K_{\epsilon}$ is a (isotropic) normal kernel. 
  
   \end{itemize}
   
   In all experiments, we use 70\% of the data for training, 15\% for validation and 15\% for testing. The exception is 
    the ZIP code example where we, for the sake of comparison, test the methods on the same 2007 images as in other works \citep{Hastie:EtAl:09}. We then use 70\% of the remaining
   images for training and 30\% for validation.\\



\noindent {\bf Evaluating the Estimators.} 
For model assessment, we compute the loss $\widehat{L}(\widehat{f},f)$ in Eq.~(\ref{lossEmpirical}) using the test data. By bootstrap, we estimate the standard error of  $\widehat{L}(\widehat{f},f)$ according to
$\sqrt{\V\left[\widehat{L}(\widehat{f},f)\right]} \approx \sqrt{\frac{1}{B}\sum_{b=1}^B \left(\widehat{L}_b(\widehat{f},f)-\overline{\widehat{L}(\widehat{f},f)} \right)^2},$
where $B=500$ is the number of bootstrap samples of the test set, $\widehat{L}_b(\widehat{f},f)$ is the estimated loss for the $b$th bootstrap sample, and
$\overline{\widehat{L}(\widehat{f},f)}$ is the mean of 
$\{\widehat{L}_b(\widehat{f},f)\}_{b=1}^B$.
 In addition to the loss (\ref{lossEmpirical}), we also perform a {\em goodness-of-fit test} to find out how well the final density estimates actually fit the observations: For every point $i$ in the test set, let 
 $U_i=\widehat{F}_{z|\vec{x}_i}(Z_i).$
 If the data are indeed distributed according to $\widehat{F}_{z|\vec{x}}$, then 
 $U_1,\ldots,U_n \overset{\mbox{\tiny{iid}}}{\sim} Unif(0,1)$. Hence, we compute the  p-value for a Kolmogorov-Smirnoff (KS) test that compares the distribution of  $U_i$ to the uniform distribution.

\vspace{-6mm}

\subsection{Numerical Examples with Simulated Data}
\label{sec::simulated}

By simulation, we create toy versions of 3 common scenarios:

\textbf{Data on Manifold.}
Data are generated according to 
 $Z|\x \sim N(\theta(\x),0.5),$
where $\x=(x_1,\ldots,x_d)$ lie on a circle with radius one
embedded in a $d$-dimensional space, and $\theta(\x)$ is the angle corresponding to the position of $\x$. We choose the data uniformly on  the manifold; i.e.,  $\theta(\x) \sim Unif(0,2\pi)$.
\vspace{2mm}

\textbf{One Relevant Covariate.}
Let 
$Z|\x \sim N(x_1,0.5),$
where $\X=(X_1,\ldots,X_d) \sim N(\vec{0},I_d)$. Here only the first covariate influences the response (i.e., the conditional density is sparse) but there is no sparse (low-dimensional) structure in $\mathcal{X}$.

\textbf{Non-Sparse Data.}
Let $Z|\x \sim N(\overline{\x},0.5),$
where $\X=(X_1,\ldots,X_d) \sim N(\vec{0},I_d)$; that is, neither the conditional density nor the input space are sparse.
\vspace{2mm}

\begin{figure}[H]
  \centering
  \subfloat{  \includegraphics[page=1,scale=0.27]{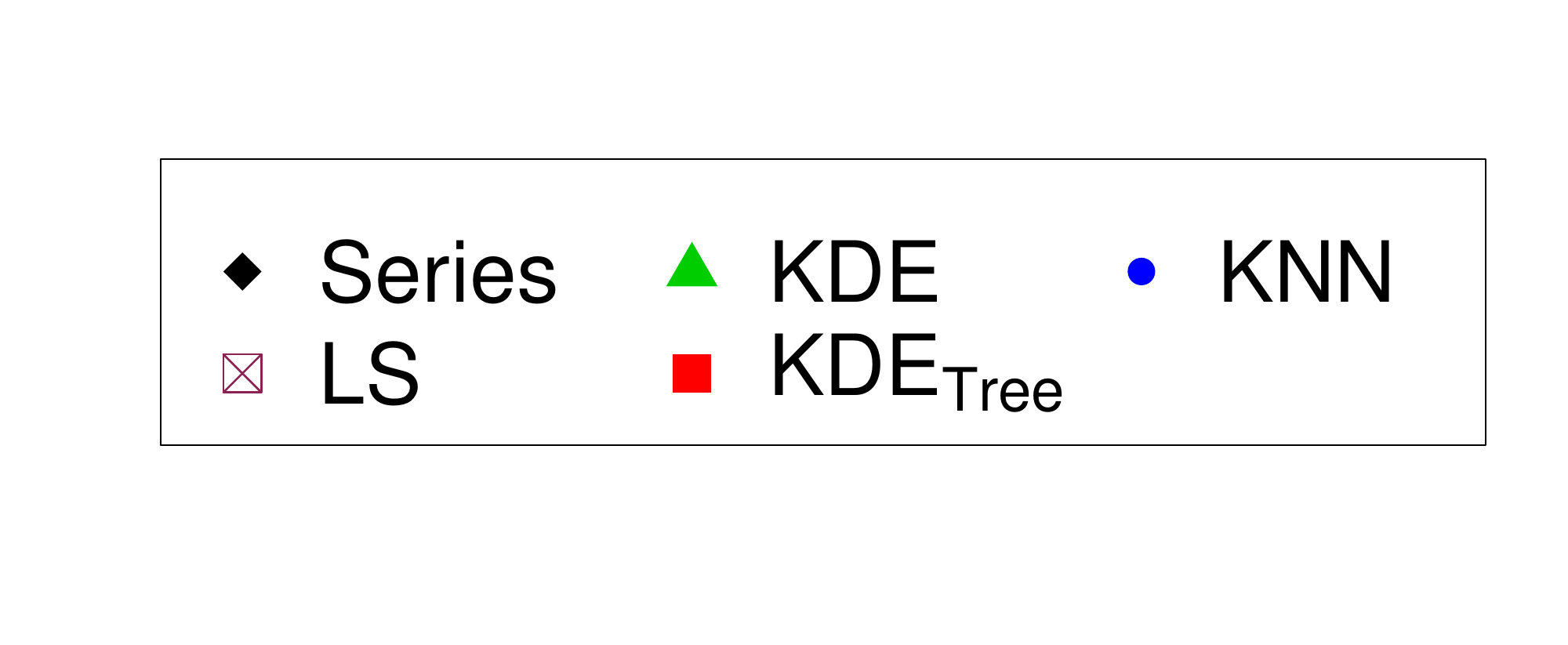}} 
   \\[-8.0mm] 
\subfloat{  \includegraphics[page=1,scale=0.27]{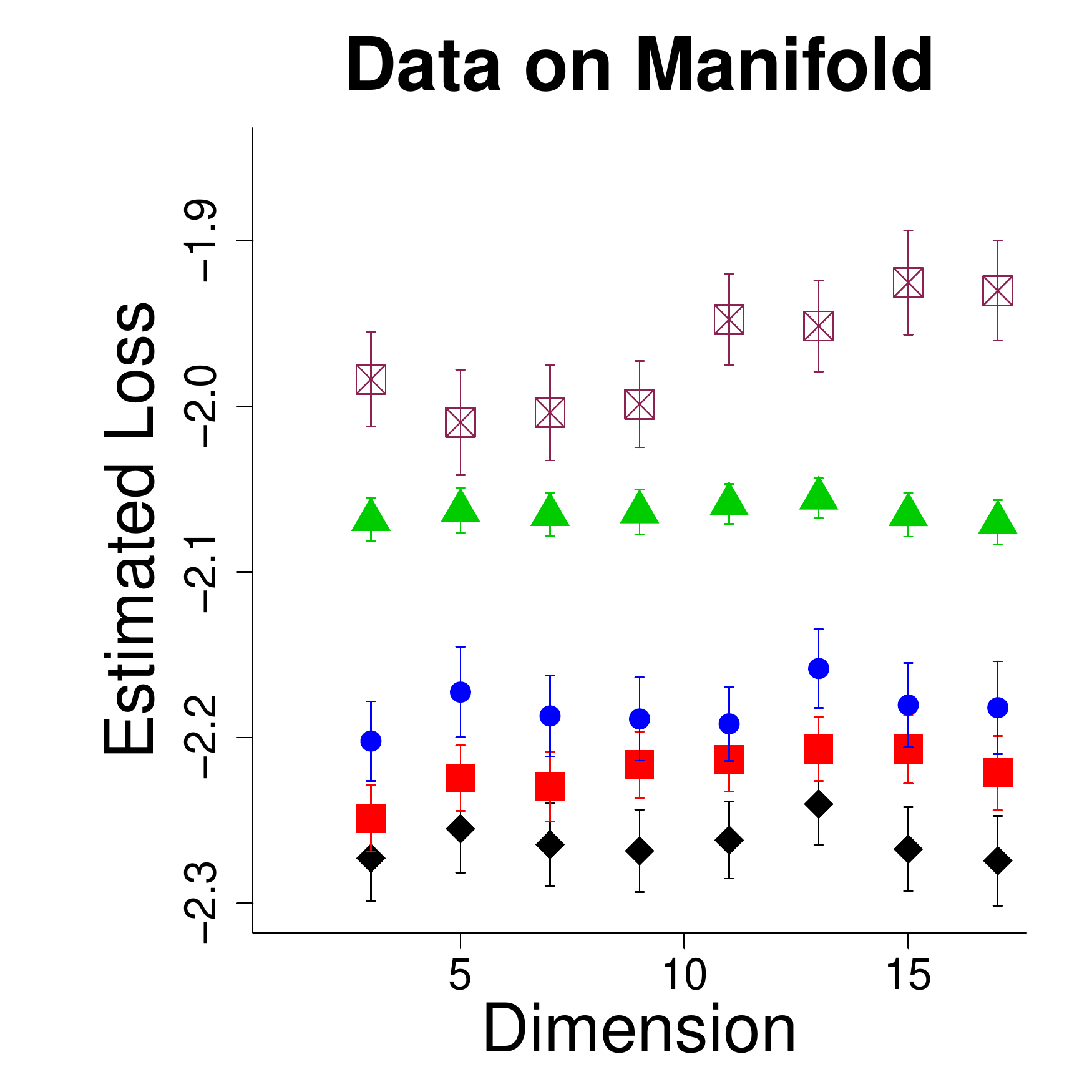}} 
\subfloat{  \includegraphics[page=3,scale=0.27]{analysisReferees.pdf}} 
\subfloat{  \includegraphics[page=1,scale=0.27]{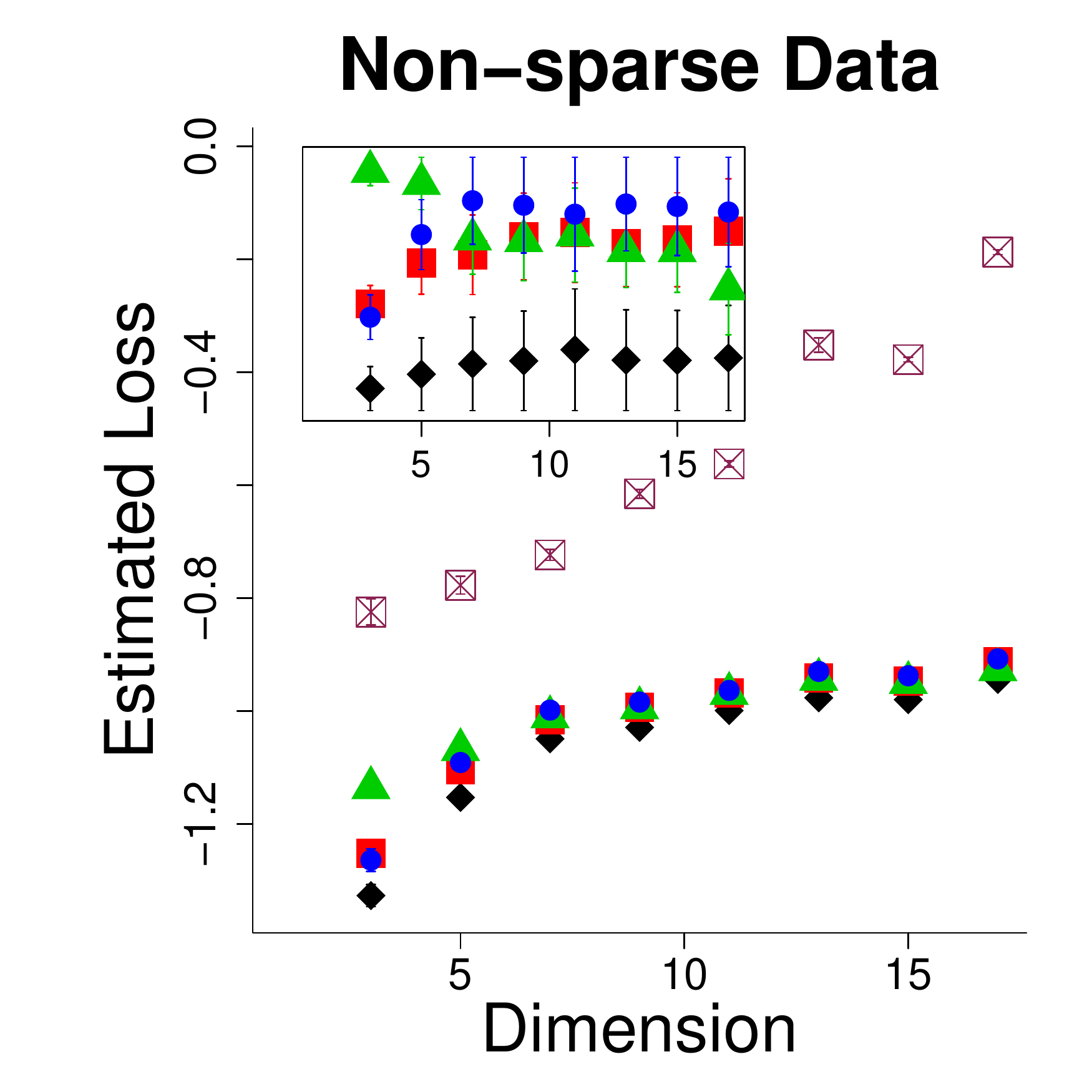}} 
  \\[-4.0mm] 
 \subfloat{  \includegraphics[page=2,scale=0.27]{analysisReferees.pdf}} 
\subfloat{  \includegraphics[page=4,scale=0.27]{analysisReferees.pdf}} 
\subfloat{  \includegraphics[page=6,scale=0.27]{analysisReferees.pdf}} 
\vspace{-3mm}
  \caption{\footnotesize Examples with simulated data. {\em Top row:} Estimated loss as a function of the dimension $d$. {\em Bottom row:} Computational time. The spectral series method (\textit{Series}) has good computational performance as a function of $d$, and it has better statistical performance than the other methods for ``Data on Manifold'' and ``Non-Sparse Data''. (The inset in the top right panel shows the loss functions after removing the \textit{LS} curve.). The online version of this figure is in color.}
  \label{fig::SimExamples}
\end{figure}

 Fig.~\ref{fig::SimExamples}  shows the estimated loss (top row) and the computational time (bottom row)  for each estimator as a function of the number of covariates $d$. For every $d$, we have repeated the simulation 200 times for $n\!=\!1,\!000$.

Our main observations are:
 \textit{KDE$_{\mbox{\tiny Tree}}$} performs well in terms of estimated loss for ``One Relevant Covariate'' (top center plot). As predicted by the theory, the statistical performance does not depend on the dimension $d$. However, 
in terms of computational time, \textit{KDE$_{\mbox{\tiny Tree}}$} becomes intractable as $d$ increases (bottom center plot): When $d=17$, each  fit takes an average of 240 seconds (4 minutes) on an Intel i7-4800MQ CPU 2.70GHz processor, compared to 24
seconds for \textit{Series}.
 For the two scenarios ``Data on Manifold'' and ``Non-Sparse Data'', \textit{Series} has the best statistical performance among the estimators. 
 Furthermore, the computational time of \textit{Series} is nearly constant as a function of the dimension $d$ in all three cases (see bottom row).
 \vspace{-4mm}
 
\begin{figure}[H]
  \centering
  \subfloat{  \includegraphics[page=1,scale=0.27]{analysisRefereesLegend.pdf}} 
   \\[-8.0mm] 
\subfloat{  \includegraphics[page=1,scale=0.27]{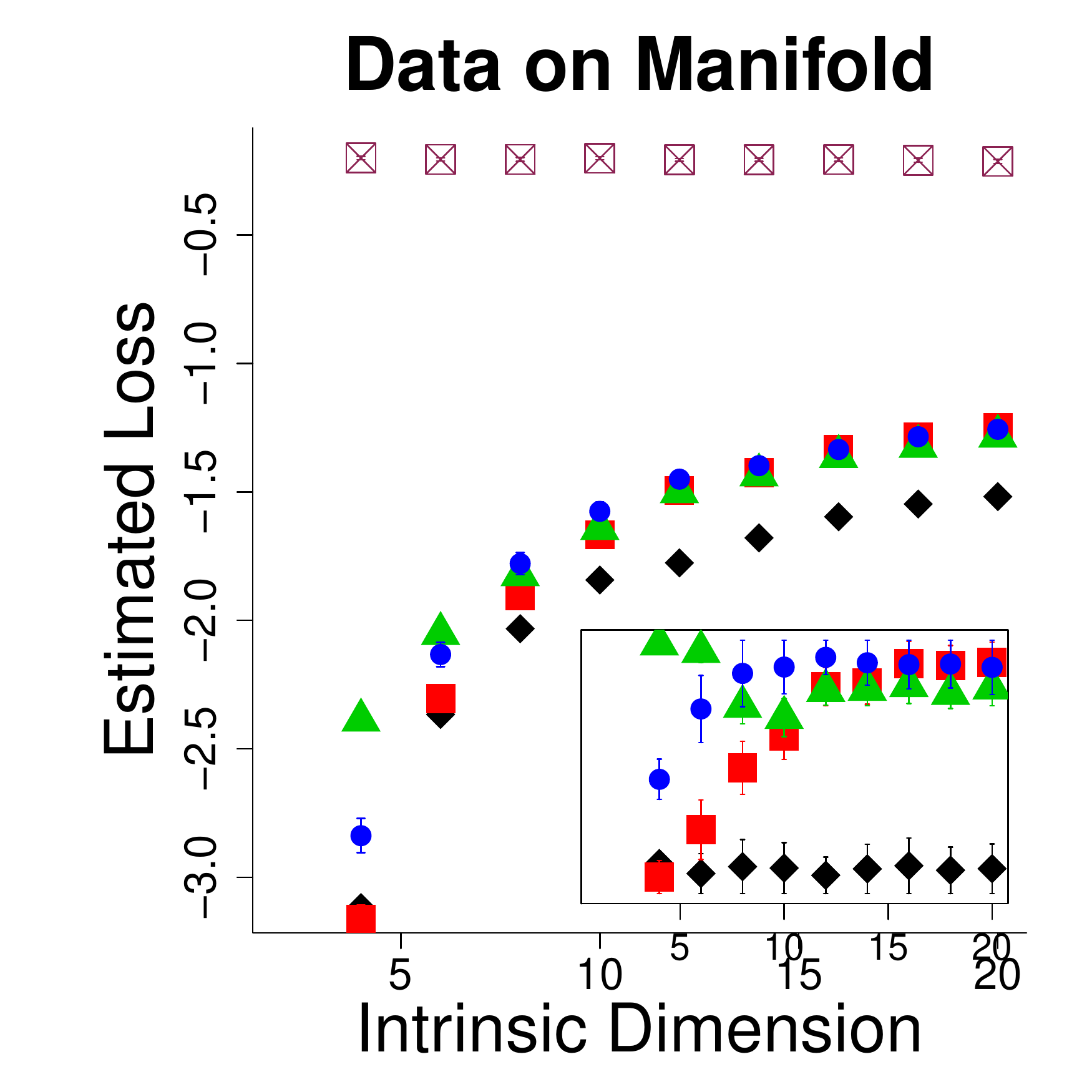}} 
\subfloat{  \includegraphics[page=1,scale=0.27]{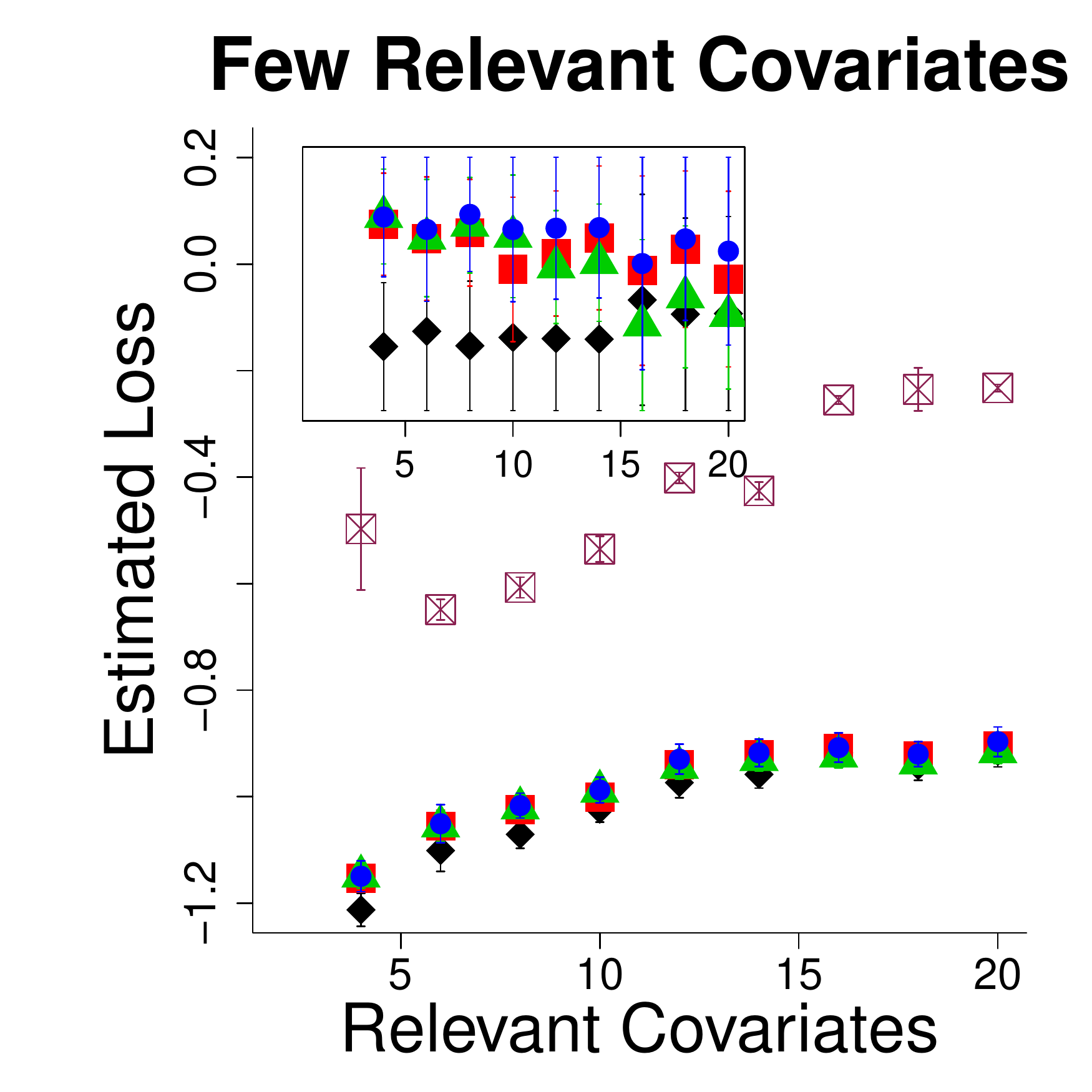}} 
  \\[-4.0mm] 
 \subfloat{  \includegraphics[page=10,scale=0.27]{analysisReferees.pdf}} 
\subfloat{  \includegraphics[page=12,scale=0.27]{analysisReferees.pdf}} 
\vspace{-3mm}
  \caption{\footnotesize Examples with simulated data. {\em Top row:} Estimated loss as a function of the intrinsic dimension (left) and
  	the number of relevant covariates (right) when the ambient dimension $d=20$. {\em Bottom row:} Computational time. The spectral series method (\textit{Series}) is computationally efficient with a better statistical performance than the other  methods. (The insets in the top panels show the loss functions after removing the \textit{LS} curves.). The online version of this figure is in color.}
  \label{fig::SimExamples2}
\end{figure}

Figure \ref{fig::SimExamples2} shows 
the results  for the scenarios ``Data on Manifold" and ``One Relevant Covariate"
when we fix the \emph{ambient} dimension at 20, and vary either the intrinsic dimension (``Data on Manifold")
or the number of relevant covariates (``Few Relevant Covariates"). See  
Supplementary Materials for more details. Contrary to what happens when 
there is only one relevant covariate (Fig.~\ref{fig::SimExamples}), when several covariates are relevant,  
\textit{KDE$_{\mbox{\tiny Tree}}$}
has similar statistical performance to \textit{Series}.
Furthermore, the computational time of \textit{Series} is nearly constant as a function of the intrinsic dimension and the number of
relevant covariates, whereas this is not the case for \textit{KDE$_{\mbox{\tiny Tree}}$}
(see bottom row).
 \vspace{-6mm}
 
Our results indicate that the series method has good statistical as well as computational performance under a variety of sparse and non-sparse settings. In the next examples, we will consider settings with large $d$.

\subsection{ZIP Code Data}
\label{ex-zip}

 Here the data are $16 \times 16$ images of handwritten digits of $\{0,1,\ldots,9\}$ from the ZIP code database from USPS \citep{Hastie:EtAl:09}. We represent each image by a vector of covariates, $\vec{x}\in \mathbb{R}^{256}$.
In addition, we define a continuous-valued response
 $Z$ according to 
 $Z|\vec{X}=\vec{x} \sim \mbox{Unif}\left(\mbox{d}(\vec{x})-\frac{1}{2},\mbox{d}(\vec{x})+\frac{1}{2}\right),$
where $\mbox{d}(\vec{x})$ is the label (i.e., the ``true'' digit associated with the image $\vec{x}$)  provided by human annotators. 

An advantage with the series estimator 
is that one, by construction, can use any orthogonal  basis to model the shape of the density $f(z|\x)$ as a function of $z$. To capture the {\em discrete} nature of the response in this example, we define an indicator basis $( \phi_i)_i$: 
 $\phi_i(z)=\I\left(z-\frac{1}{2}<i\leq z+\frac{1}{2}\right)$.
Alternatively, one could choose Haar wavelets \citep{Mallat:2009}.

\begin{figure}[H]
  \centering
\subfloat[]{\hspace{-9mm}  \includegraphics[page=1,scale=0.95]{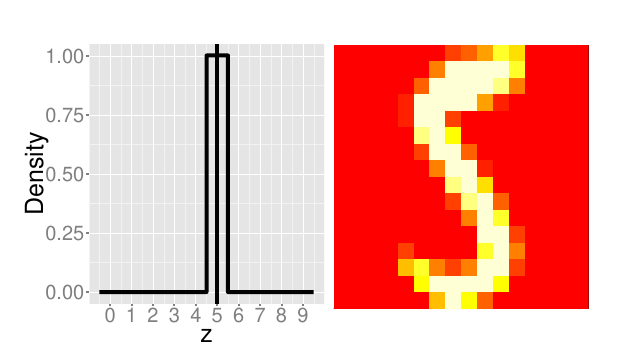}\hspace{-6mm}} 
\subfloat[]{  \includegraphics[page=1,scale=0.95]{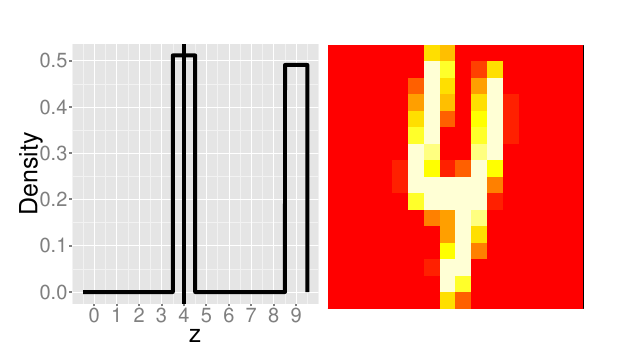}\hspace{-6mm}} 
\subfloat[]{  \includegraphics[page=1,scale=0.95]{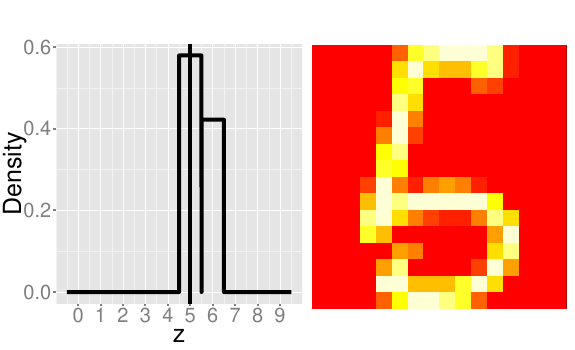}} 
\vspace{-3mm}
  \caption{\footnotesize ZIP code data from Example \ref{ex-zip}. Estimated conditional densities of the response $Z$ for 3 samples with covariates $\x$ chosen at random
  from the test data. Vertical lines indicate the imaged digit.
  The estimated densities are consistent with the images, and are multimodal when the images are atypical or ambiguous.
  Although the covariate space has $d=256$ covariates, the spectral series estimator returns reasonable estimates of $f(z|\vec{x})$. }
  \label{fig-digitsCont}
\end{figure}

Tab.~\ref{tab::results}, top row, lists the losses of the different estimators. 
The best performance is achieved by the spectral series estimator with the proposed indicator basis; 
although, \textit{Series} and  \textit{Series$_{\mbox{\tiny Diff}}$} with a standard Fourier basis already improve upon traditional methods.
Fig.~\ref{fig-digitsCont} presents density estimates $\widehat{f}(z|\vec{x})$ for 3 images. For $\approx 94\%$ of the 
 images in the test set, the estimates are unimodal and centered at the true label; image (a) is an example.
When the  estimates are {\em multimodal}, the hand-written images are atypical or ambiguous with multiple reasonable interpretations.
For example, image (b) presents characteristics of both the digit ``4" and ``9". This ambiguity is reflected in 
the estimated density which represents a mixture of two uniform distributions. The same phenomenon  
can be observed in image (c). 

Finally, although our estimator is not optimized 
for classification (which, for example, should use a 0-1 loss), one can derive a Bayes classifier from the conditional density estimates.
For  \textit{Series} with the indicator basis, this yields a classification accuracy of $94.62\%$ $(\pm 1.00\%)$, which is 
 competitive with state-of-the-art classifiers  (see e.g., \citealp{Hastie:EtAl:09}).


\subsection{Galaxy Spectra}
\label{ex-spectra}


Astronomers use redshift to determine the distances and ages of objects in the Universe.
 Typically, it is predicted from low-resolution photometric data (as in Sec.~\ref{ex-sdss}) or high-resolution spectra as in the example in this section. Here we consider the problem of estimating the redshift ($z$) of a galaxy in the  Sloan Digital Sky Survey (SDSS) using the {\em entire} spectrum ($\x$) of the galaxy. The covariates $\x$ are the flux measurements at 3501 different wavelengths; that is, the dimension $d=3501$.  Our sample consists of 2812 such spectra from SDSS DR6, preprocessed 
according to the cuts described in~\cite{Richards:EtAl:2009}. 

Because spectroscopy determines redshift with great precision, the density $f(z|\x)$ is typically degenerate, i.e., it is typically a point mass at the true redshift.
Hence, for the purpose of comparing methods, 
we add noise to the true redshift 
and let 
$z_i= z_i^{\rm SDSS}+\epsilon_i,$
where $\epsilon_i$ are $i.i.d.$ $N(0,0.02)$ and $z_i^{\rm SDSS}$ is the ``true'' redshift of galaxy $i$ provided by SDSS. 
 In other words, the conditional density $f(z_i|\x_i)$ is effectively a Gaussian distribution with mean $z_i^{\rm SDSS}$ and variance 0.02.

Tab.~\ref{tab::results} lists the results of the different conditional density estimators. Series and Series$_{\mbox{\tiny Diff}}$ clearly perform the best in terms of estimated loss. 
In addition, comparisons of the estimated and true densities together with the p-value of 0.874 for the KS test confirm that the density estimates are reasonable.  

\comment{
\begin{figure}[H]
  \centering
\subfloat{  \includegraphics[page=6,scale=0.20]{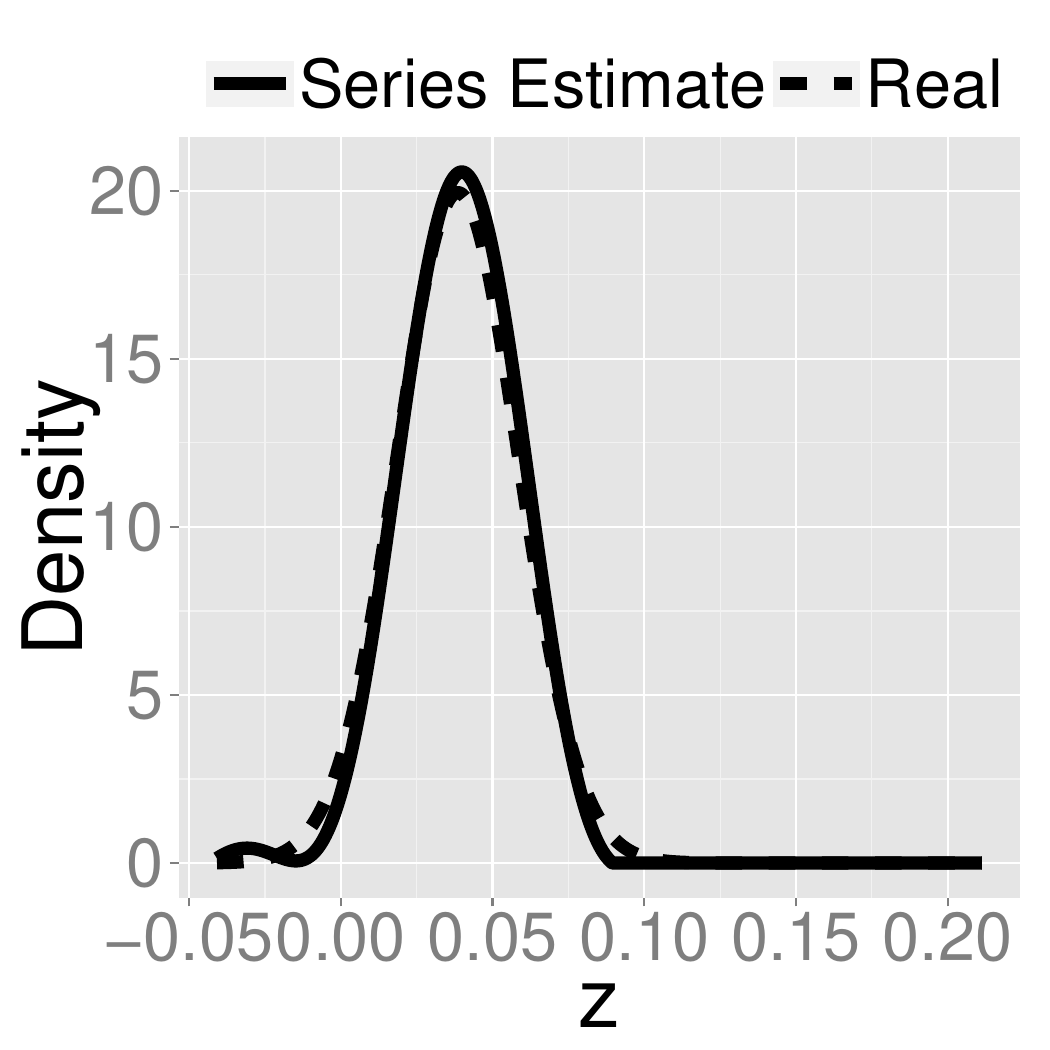}} 
\subfloat{  \includegraphics[page=20,scale=0.20]{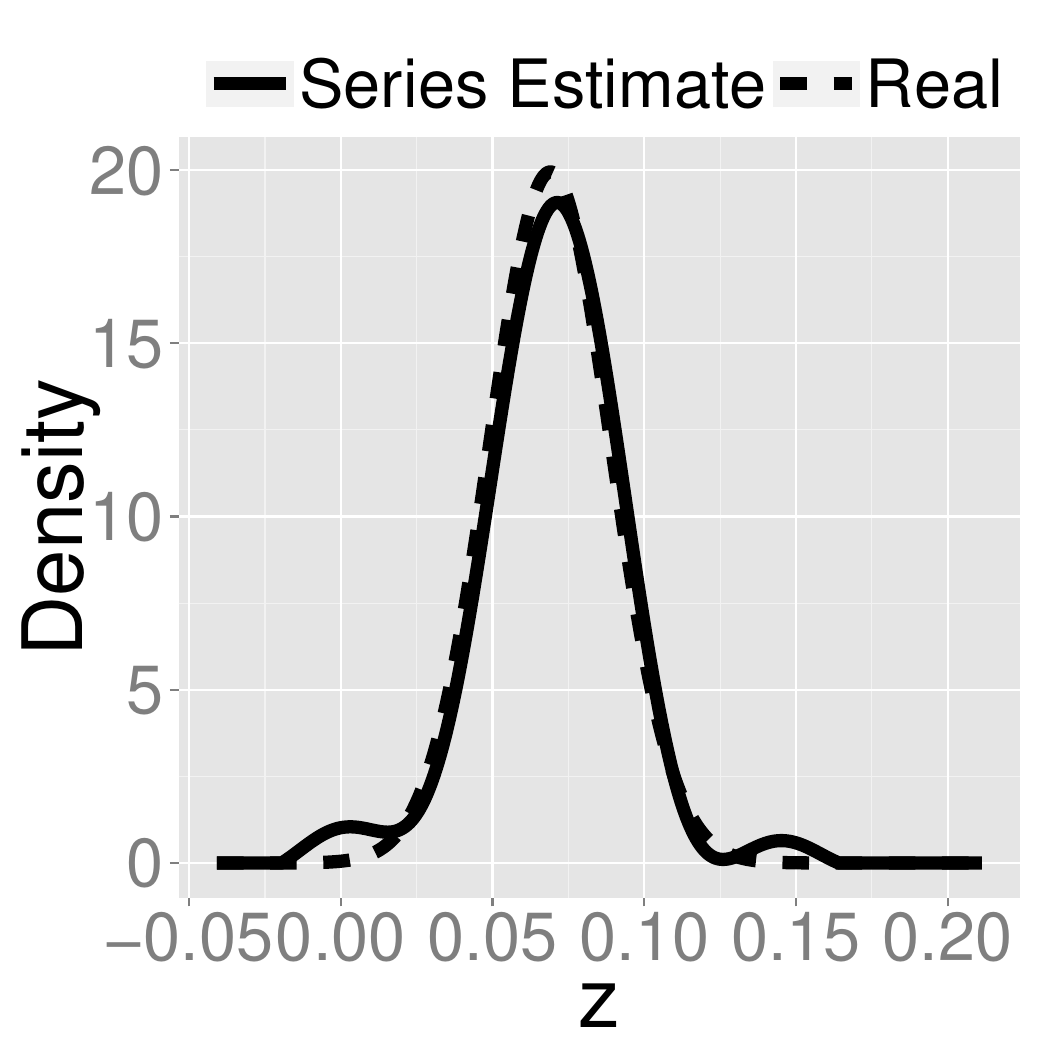}}
\subfloat{  \includegraphics[page=17,scale=0.20]{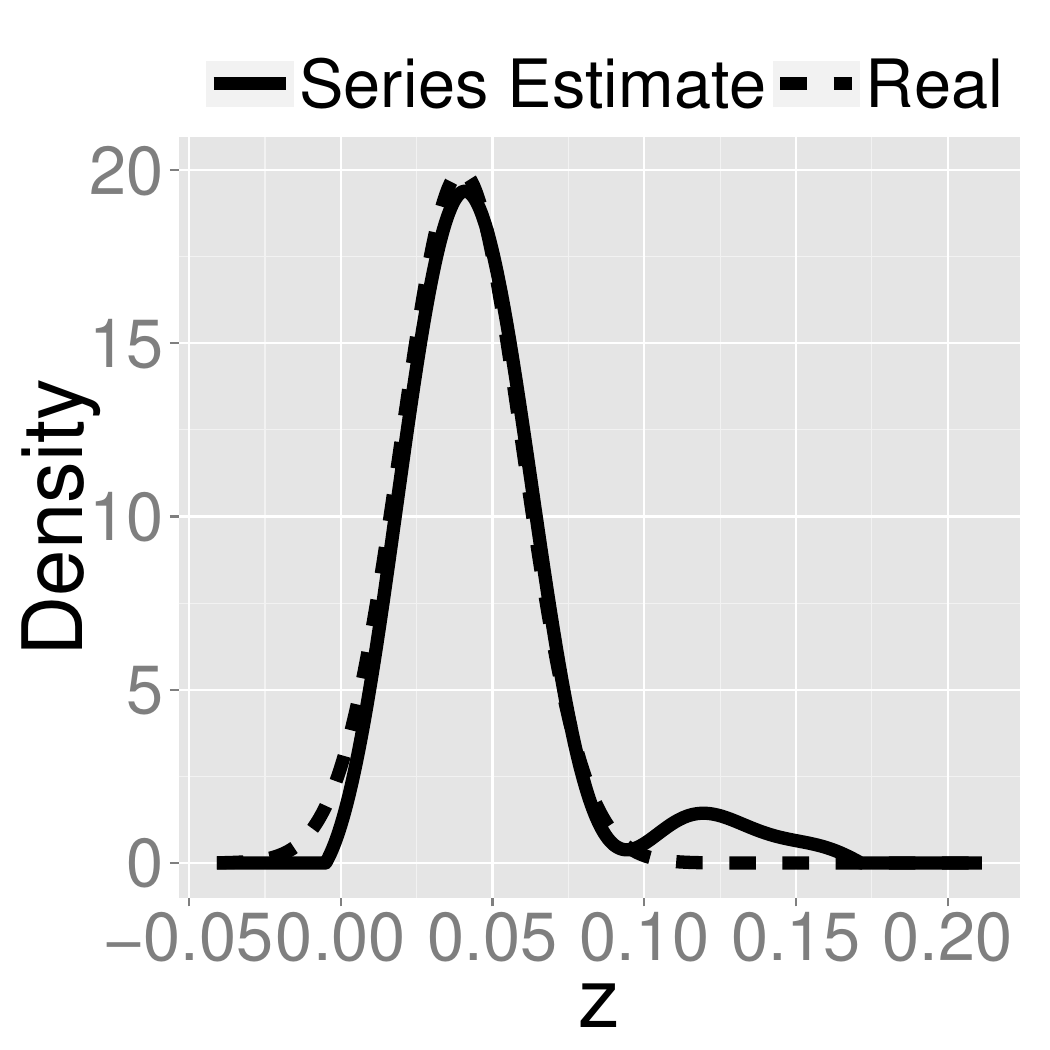}}
\vspace{-3mm}
  \caption{\footnotesize Galaxy spectra from Sec.~\ref{ex-spectra}: 
  estimated and true
  conditional densities of the response $Z$ for 3 test samples with covariates $\x$ chosen at random
    from the test data (right). 
  Even though the dimension $d=3501$, the spectral series estimator performs well.}
  \label{fig::spec}
\end{figure}
}

\subsection{Photometric Redshift Estimation}
\label{ex-sdss}

Our main application is photometric redshift estimation.
Spectroscopy allows one to estimate the redshift 
$z$ with high accuracy, but resource considerations motivate {\em photometry} ---  a measuring technique, where the 
radiation from an astronomical object is recorded via broadband filters.  
More than 99 percent of all galaxy observations are conducted via photometry. 
In photometric redshift estimation, the goal is to estimate the conditional density $f(z|\vec{x})$, where $\vec{x}$ represents the observed photometric covariates of a given object. Typically, one uses spectroscopically confirmed redshifts to train a model.  We test our CDE methods on three different sets of galaxies. In brief 
(see Appendix A.2 for details): (i) $n=3,\! 000$
 {\em luminous red galaxies} (LRGs) from SDSS with $d=12$ covariates after preprocessing~\citep{Freeman}, (ii)
  $n=10,\!  000$ galaxies  from {\em multiple surveys} with $d=10$ derived covariates~\citep{Sheldon}, and (iii) $n=752$  galaxies from {\em COSMOS} (T.~Dahlen 2013, private communication) with $d=37$ covariates derived from a variety of photometric bands.
  
The bottom three rows of Tab.~\ref{tab::results} summarize the results 
of the different conditional density estimators. As in previous examples, the two spectral series estimators perform the best, followed by the KNN.
In terms of loss, the advantage of spectral series is most apparent for the COSMOS data; this is the most challenging data set as the number of covariates (37)
is large compared to the training sample size.
The conditional density estimates are reasonable, but  there is still room for improvement 
for COSMOS. The KS test returns a p-value of 0.045 for these data, in contrast to 0.393 for  {\em luminous red galaxies} and 0.071 for {\em multiple surveys} data.

\begin{figure}[H]
  \centering
 \subfloat{  \includegraphics[page=14,scale=0.18]{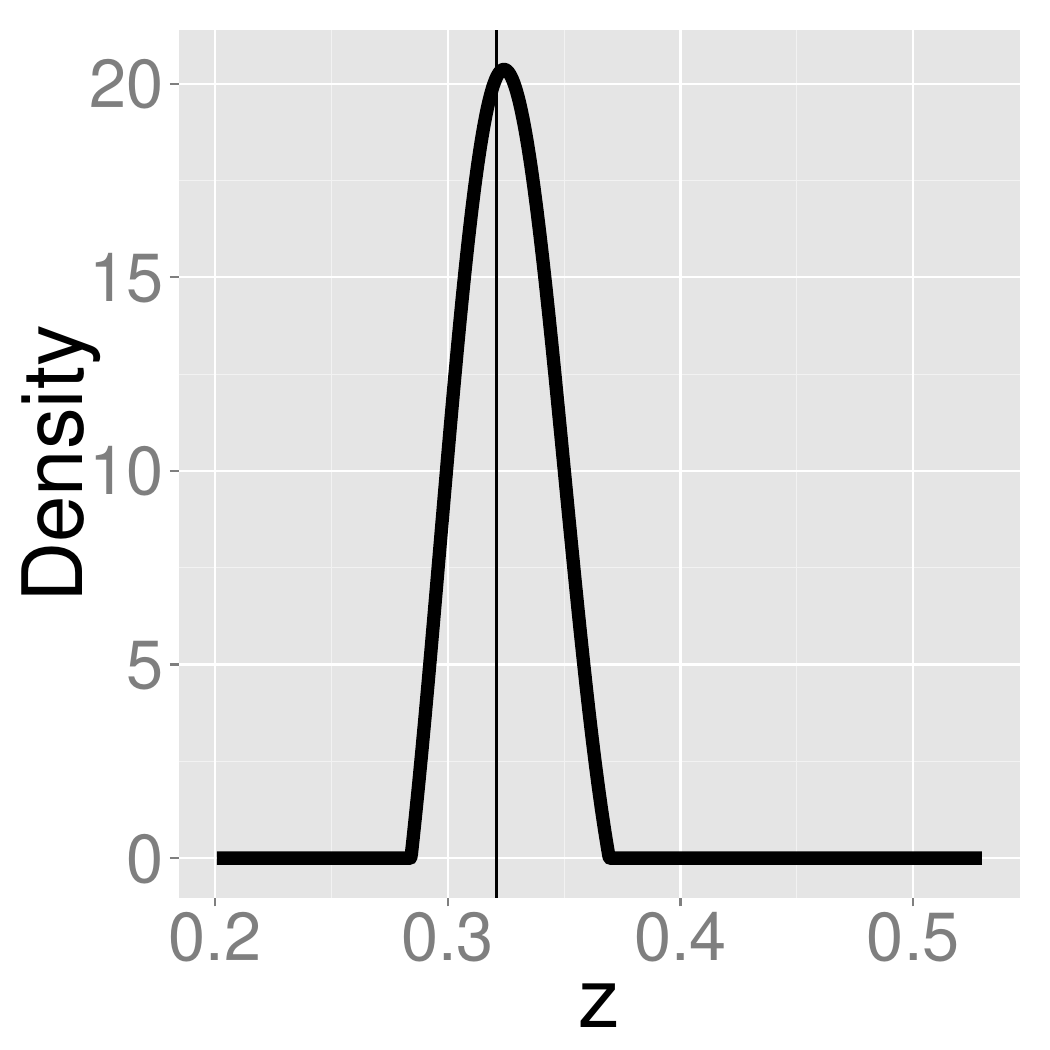}} 
 \subfloat{  \includegraphics[page=17,scale=0.18]{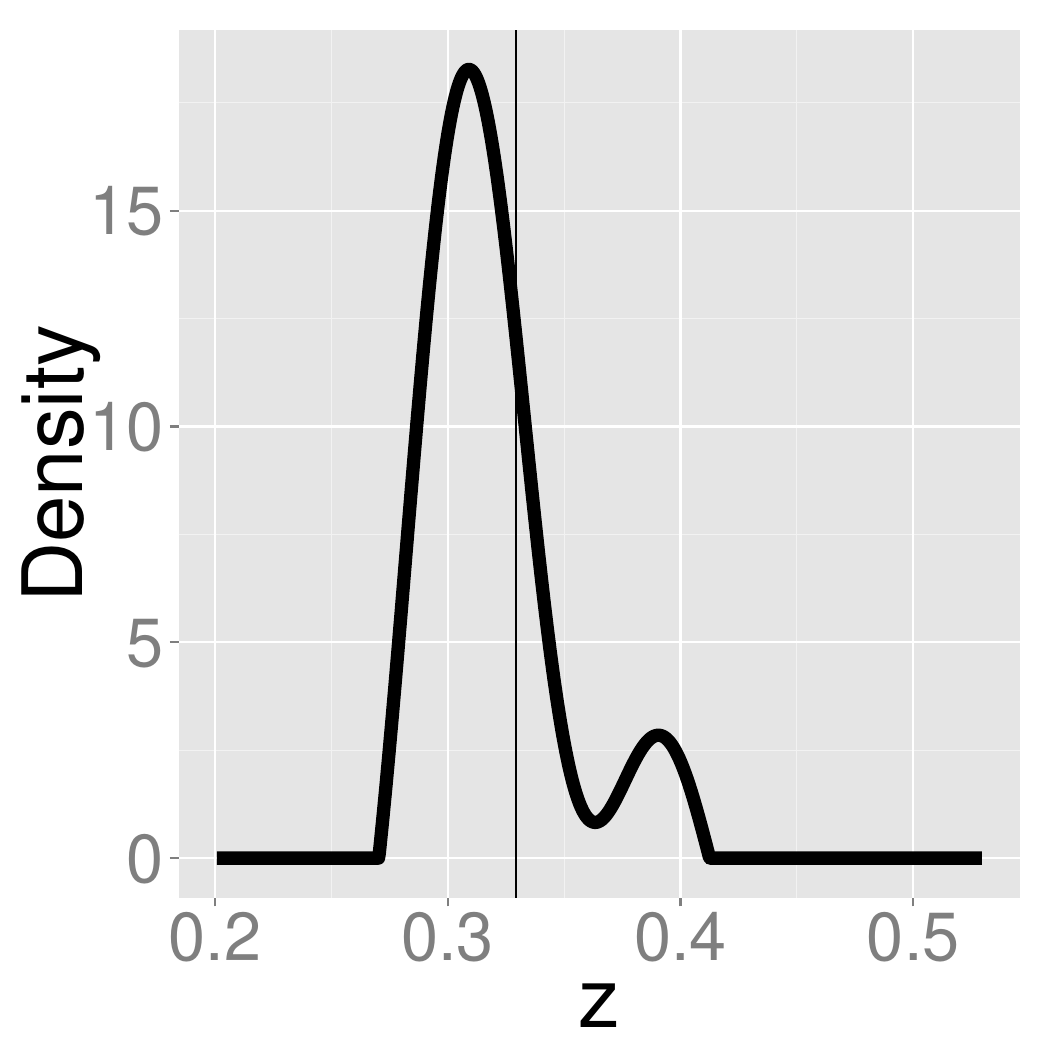}}
 \subfloat{  \includegraphics[page=6,scale=0.18]{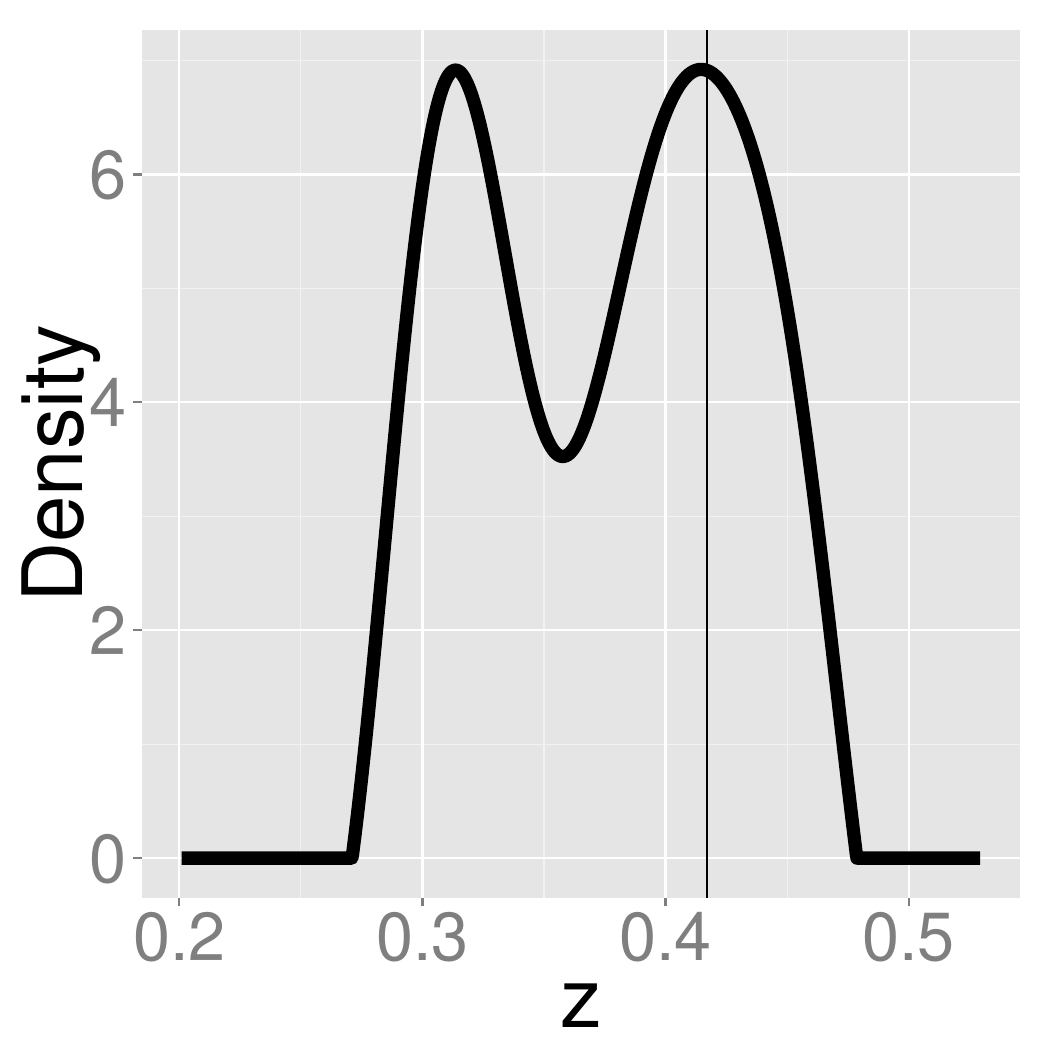}}
 \subfloat{  \includegraphics[page=38,scale=0.18]{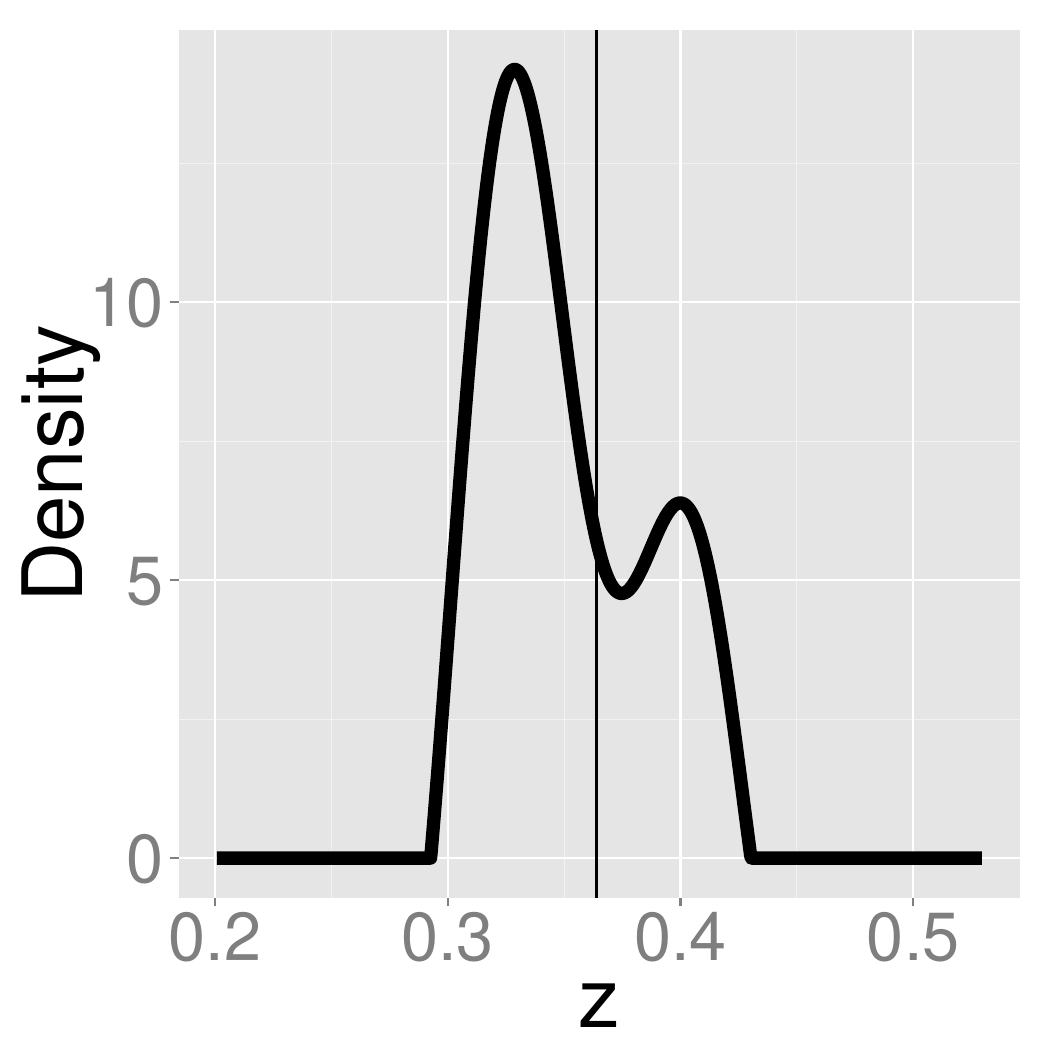}}
\vspace{-3mm}
  \caption{\footnotesize Estimated densities for 4 (randomly chosen) luminous red galaxies from the SDSS test set. 
  Vertical lines indicate spectroscopically observed redshift values.}
  \label{fig::photoZ}
\end{figure}

Fig.~\ref{fig::photoZ} shows examples of spectral series density estimates for galaxies in SDSS. 
The multimodal and asymmetric densities are particularly informative to astronomers. Typically, they correspond to cases where a single-point estimate (e.g., the regression $\E[Z|\vec{x}]$, or the mode of $f(z|\x)$) may induce large errors in cosmological analyses. 

\begin{figure}[H]
  \centering
\subfloat{  \includegraphics[page=2,scale=0.21]{photoZComputationalTimeIncreasingN_3.pdf}}
\subfloat{  \includegraphics[page=1,scale=0.21]{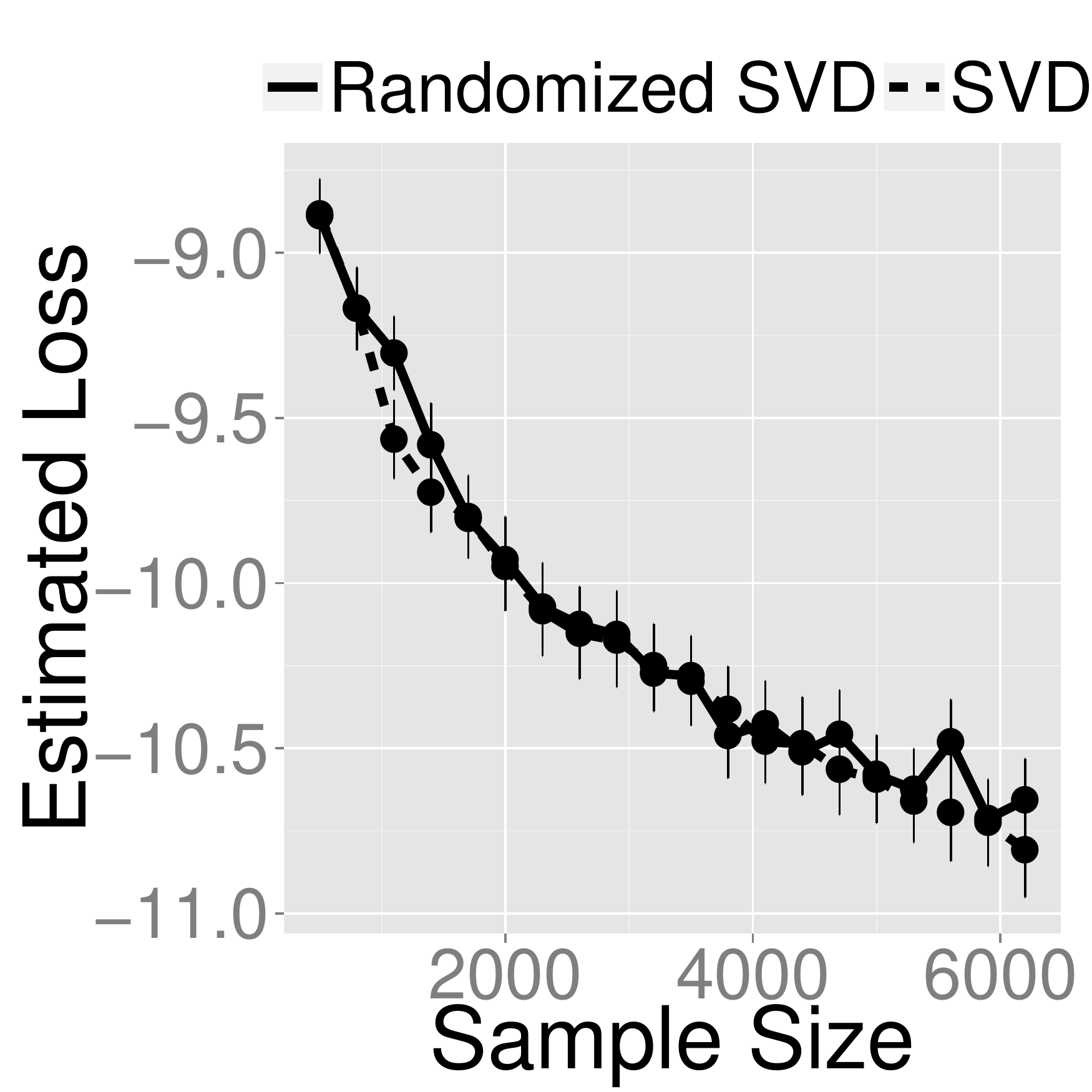}}\hspace{23mm}
\subfloat{  \includegraphics[page=2,scale=0.20]{photoZSparseMatrixScaleFinal_2.pdf}}
  \caption{\footnotesize {\em Left panel:} Randomized SVD  can dramatically reduce the computational time for large sample sizes (left plot) with almost no 
  loss in statistical performance (center plot). {\em Right panel:} With sparse Gram matrices, one can 
  cut down the memory use with about 30\% with little loss in statistical performance. }
  \label{fig::photo_time}
\end{figure}

{\bf Scalability.} 
 Fig.~\ref{fig::photo_time} indicates massive 
 payoffs in implementing Randomized SVD. 
Even without parallelization, we are able to cut down the computational time with a factor of $5$ (left plot) with almost no decrease in statistical performance (center plot). (In the experiments, we use the data by \citet{Sheldon} and vary the size of the training set for a fixed number of 3,000 validation samples and 10,000 testing samples.)
Similarly, we can save  30\% of the memory with little loss in statistical performance by thresholding the Gram matrix (right plot). 
(Here we vary the threshold $\xi$ in Sec.~\ref{sec::improv} for 5,000 training, 2,500 validation and 2,500 test examples.) 

\begin{table}[H]
\caption{\footnotesize  Estimated $L^2$ loss (with standard errors) in conditional density estimation. Best-performing models with the smallest loss are in bold fonts. The \dag-symbol denotes results with the indicator basis.
 Note that {\it KDE$_{\mbox{\tiny Tree}}$} cannot be applied to ``ZIP Code'' and ``Spectra'' due to the method's high computational cost in high dimensions.} \label{table::loss}
\vspace{1mm}
\centering
\tabcolsep=0.1cm
{\footnotesize
\begin{tabular}{l|c|cccccc}
{\bf Data Set}  & {\bf Dim}  &\multicolumn{6}{c}{{\bf Loss}}       \\
 & & {\it Series }& {\it Series$_{\mbox{\tiny Diff}}$} &   {\it LS}  & {\it KDE} & {\it KDE$_{\mbox{\tiny Tree}}$} & {\it KNN} \\
 \hline 
\multirow{2}{*}{ZIP Code }& \multirow{2}{*}{256}&  -3.94 (0.09)   &  -3.84  (0.09)  &  
\multirow{2}{*}{-0.15 (0.06)} & \multirow{2}{*}{-3.34 (0.05)} & \multirow{2}{*}{---}& \multirow{2}{*}{-3.60 (0.10)}\\[-2mm]
 & &\textbf{-4.47 (0.08)}\dag  &  \textbf{-4.42  (0.10)}\dag & & &  \\ \cline{3-8}
 
Spectra             & 3501&\textbf{-1.75 (0.06)} &\textbf{-1.77 (0.07)}     & -0.26 (0.02)     &-1.20 (0.05) & --- & -1.61 (0.07) \\ \cline{3-8}
Photo-z LRGs & 12&\textbf{-1.88 (0.07)} &\textbf{-1.84 (0.06)}       & -1.53 (0.05)    &-1.72 (0.06) & -1.56 (0.04) & -1.72 (0.07) \\ \cline{3-8}
Photo-z Multiple & 10&\textbf{-11.81 (0.20)} &\textbf{-11.49 (0.21)}  & -8.49 (0.25)  & -9.40 (0.19) &-7.04 (0.09) & --11.06 (0.21)  \\ \cline{3-8}
Photo-z COSMOS & 37& \textbf{-9.49 (1.03)} &\textbf{-9.02 (0.97)}   &  -0.23 (0.02) & -5.59 (1.39) &-0.60 (0.01) &-6.98 (0.88) \\
\end{tabular}
\label{tab::results}
}
\end{table}

%
%

\section{Conclusions}
\label{sec-conclusions}

{\em Orthogonal series estimation} is a classical approach to nonparametric inference but has so far been limited to less than 10 covariates.
For the first time in the literature, we present theoretical and empirical evidence that orthogonal series methods -- with the right choice of basis -- can be effective in dimensions with upwards of $10^3$ variables. 
 Our series approach to conditional density estimation is data-driven and has the advantage of a fast implementation with only one tensor product. The method directly expands the conditional density $f(z|\vec{x})$ in   eigenfunctions that adapt to the geometry of the data and does not require dividing two density estimates, or estimating $f(\vec{x})$, both difficult tasks in higher dimensions. Although one has to estimate the basis $\Psi$, our rate calculations show that if $f(z|\vec{x})$ is smooth with respect to $\Psi$, one still benefits when compared with estimation methods that do not take the geometry
of the data into account -- especially when the dimension $d$ of the data is large. This result is confirmed by our experiments.

\comment{ The spectral series estimator naturally fits a semi-supervised learning framework, where one has access to both unlabeled and labeled data --- a common scenario in modern applications where the data collection itself may be cheap but labeling the data (i.e., providing the response $z$) may be labor intensive. If the data lie on a submanifold with intrinsic dimension $p$, then in the limit 
 of infinite unlabeled data, 
 the estimator adapts to the geometry of the data and achieves the minimax rate of nonparametric estimators {\em in $p$+1 dimensions}. On the other hand, if there is no unlabeled data (as in the numerical examples in this paper), we guarantee
$O_P\left(n^{-1/O\left(p^2\right)}\right)$ rates which for $p \ll d$ is still considerably better than the standard minimax rate for estimating functions in $d+1$ dimensions.} 

There are also benefits to explicitly computing the eigenvectors of a kernel: The eigenvectors provide coordinates for the data and allow the data analyst to visualize and explore complex high-dimensional data, functional data, and abstract objects in a graph. 
   By introducing an orthogonal series approach to high-dimensional inference, we open up the doors to a whole range of possibilities of using  Fourier series and spectral bases for statistical analysis of complex data. Future work includes adapting the method to massive data by implementing approximate nearest neighbor searches and randomized eigendecompositions via multi-processor architectures. In addition, in a separate paper, we will investigate the use of spectral series for estimating other unknown functions $g: \mathcal{X} \rightarrow \mathbb{R}$ for high-dimensional aggregate objects $\x \in \mathcal{X} \subset \mathbb{R}^d$ with complicated dependence structure. In particular, we will estimate density ratios $\beta(\x)=f(\x)/g(\x)$ and the likelihood function $\mathcal{L}(\x;\theta)$ of observing complex data $\x \in \mathcal{X}$ given parameters $\theta$.\\

{\small \noindent \textbf{Acknowledgments.}
We thank Peter E. Freeman, Jing Lei and Chad M. Schafer
 for their insightful comments.  We are also grateful to the referees and associated editor for all the detailed comments that helped improve the paper.
 This work was partially supported by \emph{Conselho Nacional de Desenvolvimento Cient\'ifico e Tecnol\'ogico} (200959/2010-7),
 \emph{Funda\c{c}\~ao de Amparo \`a Pesquisa do Estado de S\~ao Paulo} (2014/25302-2), the \emph{Estella Loomis McCandless Professorship}, and NSF DMS-1520786.}\\


{\small \noindent \textbf{Supplementary Materials.} The following files are available online: 
 
 \begin{itemize}
  \item   {\tt Appendix.pdf} (PDF file)
with  
 numerical examples, theoretical analyses, proofs, details on the simulated examples from Fig.~\ref{fig::SimExamples2} and details on the galaxy data in Sec.~\ref{ex-sdss}. 

 \item {\tt specSeriesCDE.tar.gz} (Compressed tar file) with 
 R code for the spectral series estimator.
 
\item  {\tt codeAndData.tar.gz} (Compressed tar file)
with the R code and data used in the examples.

 \end{itemize}}

{ \footnotesize 
\bibliography{JCGSArxiv}
}

\end{document}